%% file: 00flow.tex
\documentstyle[epsf,12pt]{article}
\begin{document}
\title{Light-cone Hamiltonian flow for positronium}
\author{Elena L. Gubankova$^{1}$\thanks{Present address: Department of
Physics, North Carolina State University, Raleigh, 27695-8202 NC USA},  
 Hans-Christian Pauli$^{2}$,
 Franz J. Wegner$^{1}$\\
and G\'abor Papp$^{1,3}$\thanks{on leave from 
 Inst. for Theor. Physics, E\"otv\"os University, P\'azm\'any P.s. 1/A,
 Budapest, H-1117}\\
$^1$ Institut f\"ur Theoretische Physik der
Universit\"at Heidelberg,\\
 D-69120 Heidelberg \\
$^2$ Max-Planck-Institut f\"ur Kernphysik, D-69029 Heidelberg \\
$^3$ CNR, Department of Physics, Kent State University,\\ 
     44242 OH USA  
}
\date{Fri, 29 Jan 1999}
\maketitle
\begin{abstract}
The technique of Hamiltonian flow equations is applied to 
the canonical Hamiltonian of quantum electrodynamics
in the front form and 3+1 dimensions. 

The aim is to generate a bound state equation
in a quantum field theory, 
particularly to derive an effective Hamiltonian which is
practically solvable in Fock-spaces with reduced particle number.
The effective Hamiltonian, obtained as a solution of flow eqautions
to the second order, is solved numerically for positronium spectrum.
The impact of different similarity functions is explicitly studied.

The approach discussed can ultimately be used to address to the
same problem for quantum chromodynamics.
\end{abstract}
\vfill
\noindent
Preprint: MPI-H-V33-1998\\
PACS-numbers are:\\
  {11.10.Ef}\quad {Lagrangian and Hamiltonian approach}\\ 
  {11.15.Tk}\quad {Other non-perturbative techniques}\\ 
  {12.38.Aw}\quad {General properties of QCD (dynamics, confinement, etc.)}\\
  {12.38.Lg}\quad {Other non-perturbative calculations}
\hfill
\newpage\tableofcontents\newpage
\section{Introduction}
\label{sec:1}

Over the past twenty years two fundamentally different pictures
of hadrons have developed. 
One, the constituent quark model is closely related to 
experimental observation and phenomenology.  
The other,  quantum chromodynamics (QCD) is based on a
covariant non-abelian quantum field theory. 
Disregarding lattice gauge calculations, one has several reasons 
\cite{wil89} why the front form of Hamiltonian 
dynamics \cite{dir49}, as reviewed recently in \cite{bpp97}, is one of 
the very few candidates for reconciling the two approaches. 
Particularly the simple vacuum and the simple boost properties
confront with the complicated vacuum and the complicated 
boosts in the conventional Hamiltonian theory.
Wilson and collaborators \cite{GlWi,WiPe} have proposed a scheme
in which one presumes a potential for the bound-states
and handles the relativistic effects by structures imposed  
by the needs of renormalization.
The available numerical examples \cite{BrPe,JoPeGl} however
violate admittedly some symmetries of the Lagrangian
and it is not clear how to restore them systematically.

There are two major problems when one addresses to solve a 
Hamiltonian bound state equation 
\begin{equation}
   H \vert\psi\rangle =E \vert\psi\rangle
\label{eq:i1}\end{equation}
in a covariant relativistic field theory. 
First, the canonical field-theoretical Hamiltonian $H$ contains 
states (fields) with arbitrarily large energies.
Second, the number of particles in a field theory is unlimited
and $H$ contains the impact of arbitrarily many particles. 
An eigenfunction $\vert\psi\rangle$, for example a meson wave function, 
has contributions from arbitrarily many Fock-space sectors
$\vert\psi\rangle = \varphi_{q\bar{q}}\vert q\bar{q}\rangle
 + \varphi_{q\bar{q}g}\vert q\bar{q}g\rangle + \dots$.
Therefore, in general, the Hamiltonian operator $H$ 
can be understood as a matrix with infinite dimensions both with
respect to `energy' and with respect to `particle number'.
The method displayed below, the `Hamiltonian flow equations' 
of Wegner \cite{weg94}, cope with either of them.

It is a subject of its own and not completely trivial to write
down a suitable Hamiltonian (operator) $H$ \cite{bpp97}.
In the sequel we shall use the canonical (light-cone) Hamiltonian 
of gauge theory in the light-cone gauge ($A^+=0$).
In the light-front Hamiltonian approach one faces then two classes 
of problems that are related with each other:
the problems associated with the light-front formulation and 
the problems in formulating an effective Hamiltonian theory
and the difficulties in the renormalization program.
In order not to ponder all these problems with the
problems of confinement and chirality, one disregards in this work
QCD and restricts to QED as a model case.
Unlike in other work \cite{JoPeGl,pau96,TrPa} on QED and other models
we apply in this work the method of flow equations 
\cite{weg94,LeWe,GuWe,gub98}
with the objective to derive from the Hamiltonian in the front form
a well-founded effective  low-energy Hamiltonian which acts 
in the space of a few particles and which can be solved 
explicitly for bound states.

Since we aim at a pedagogical presentation we
sketch shortly the ingredients of front-form QED in Sec.~\ref{sec:3}.
In Sec.~\ref{sec:2} the general aspects of the Hamiltonian
flow equations are collected in such a form that the
application of the flow equations to QED in Sec.~\ref{sec:4} 
becomes more transparent. It is here where the bulk 
of the present work is displayed in a formal way.
The implications are discussed in Sec.~\ref{sec:5},
and the numerical calculations are given in Sec.~\ref{sec:6}. 
The work continues to be actively pursued, and this is only the
first in a series of papers.

\section{Flow equations and bound state problem}
\label{sec:2}
In this work, we shall focus on
the flow equations for Hamiltonians formulated first by 
Wegner \cite{weg94} aiming at the construction of an effective 
bound state Hamiltonian for field theories.
The general aspects of the method do not depend on the 
nature of the Hamiltonian and  
before plunging into the paraphernalia of the field theoretical
details it is useful to outline the general ideas 
in a manner slightly different from the original 
formulation \cite{weg94}.

It is always possible to divide the complete Fock space 
(with its different particle number sectors)
into two arbitrary subspaces, called the $P$- and the $Q$-space.
The Hamiltonian matrix of Eq.(\ref{eq:i1}) has then the form
\begin{eqnarray}
   H = \left(
   \begin{array}{cc}
      PHP & PHQ \\
      QHP & QHQ
   \end{array}\right)
\,,\label{eq:i3}\end{eqnarray}
with $P$ and $Q=1-P$ being projection operators.
Suppose now we are unable to solve the eigenvalue equation 
for the whole matrix, because of, say, computer limitations. 
The method of flow equations allows then to unitarily transform the 
Hamiltonian matrix into a block-diagonal form 
\begin{eqnarray}
   H_{\rm eff} =\left( 
   \begin{array}{cc}
     PH_{\rm eff}P & 0            \\
     0             & QH_{\rm eff}Q
   \end{array}\right)
\end{eqnarray} 
by analytical procedures. 
The two blocks are then decoupled, and the eigenvalue problem 
with an effective Hamiltonian $H_{\rm eff}$ can be solved and 
diagonalized separately for either of the two spaces, which 
technically might (or might not) be easier than the solution
of the full problem.

Since one can choose the number of particles in the $P$-space
one reduces in this way in general the many-body problem 
to a few particle bound state problem
at the expense of finding a more complicated effective Hamiltonian 
operating in a limited particle number space.
This general idea is similar to the procedure of Tamm and Dancoff 
\cite{tam45,dan50} where an effective interaction in a few particle
sector was obtained by eliminating the `virtual scatterings' 
to the higher  Fock-space sectors in the $Q$-space, 
see also \cite{pau96}.

The method of flow equations \cite{weg94,LeWe,GuWe,gub98} 
works with a unitary transform which is governed by a 
continuous parameter $l$. The unitarily transformed
Hamiltonian is then a function of this parameter, {\it i.e.}
\begin{eqnarray}
   \frac{dH}{dl} = [\eta(l), H(l)]
\,.\label{eq:i4}\end{eqnarray}
The generator of the transformation is subject to some choice
but taken here as \cite{weg94}
\begin{eqnarray}
   \eta(l) = [H_d(l), H(l)]
\,.\label{eq:i5}\end{eqnarray}
The Hamiltonian is separated conveniently into a block-diagonal
part 
\begin{eqnarray}
   H_d(l) =\left(
   \begin{array}{cc}
     PH(l)P & 0     \\             
     0      & QH(l)Q     
  \end{array}  \right)
\,,\end{eqnarray}
and into the rest 
\begin{eqnarray}
  H(l)-H_d(l)=\left( 
  \begin{array}{cc}
    0      & PH(l)Q \\
    QH(l)P & 0
  \end{array} \right)
\,,\end{eqnarray}
which is purely off-diagonal. It is precisely this rest
which due to $PHQ$ `changes the particle number'.
In many cases of practical interest one can interpret
this rest as a `residual interaction'. If the rest vanishes,
or if it is exponentially small, one has solved the most important
part of the problem. In the sequel we convene that the flow
parameter changes from $l=0$ to $l\rightarrow\infty$, corresponding
to a change from the initial canonical Hamiltonian to  
block-diagonal effective Hamiltonian.
According to Eq.(\ref{eq:i5})  the generator is always off-diagonal.
\begin{eqnarray}
   \eta(l)=\left(
   \begin{array}{cc} 
     0          & P\eta(l)Q  \\
     Q\eta(l)P  & 0
   \end{array} \right)
\,.\end{eqnarray}
The flow equations Eq.(\ref{eq:i4}) for the diagonal and the 
rest sectors can then be disentangled into
\begin{eqnarray}
   \frac{d}{dl} PHP &=& P\eta QHP - PHQ\eta P
\,,\label{eq:i8}\\
   \frac{d}{dl} PHQ &=& P\eta QHQ - PHP\eta Q
\,,\label{eq:i9}\end{eqnarray}
and the trivial identity
\begin{equation}
   P\eta Q = PHPHQ - PHQHQ
\,.\label{eq:i10}\end{equation}
For $QHQ$, $QHP$, and $Q\eta P$ one proceeds correspondingly.
Since ${\rm Tr} PHQHP$ is restricted from above,
Wegner's choice for the generator Eq.(\ref{eq:i5}) 
results in a monotonously decreasing measure 
for the off-diagonal particle-number changing interaction
\begin{eqnarray}
    \frac{d}{dl} {\rm Tr} PHQHP 
    &=& {\rm Tr} \Big( P\eta Q (QHQHP - QHPHP)\Big)
\nonumber\\
    &+& {\rm Tr} \Big((PHPHQ - PHQHQ) Q\eta P \Big) 
\nonumber\\
    &=& 2 {\rm Tr} (P\eta Q\eta P)\leq 0
\,.\label{eq:i11}\end{eqnarray}
For the generator this implies that
$\eta(l)\rightarrow 0$ in the limit $l\rightarrow\infty$
and that the block-diagonal part of the Hamiltonian commutes 
with the Hamiltonian itself.
Thus, the effective Hamiltonian $H_{\rm eff}$ becomes
`more and more block-diagonal' with increasing flow parameter.

In general the solution of these equations will become quite involved.
One reason is, that the equations are nonlinear. Another is, that starting
with a two-particle interaction one generates due to the
commutators three-particle, four-particle etc. interactions.
It is however possible to solve the equations in certain limits or 
approximations. A limit in which
the equations can be solved exactly to a large extend is the 
$n\rightarrow\infty$ limit of an $n$-orbital model \cite{weg94}. In this
limit the equations for the two-particle interaction are closed,
that is generated three-particle interactions do not couple back to
the flow equations for the leading two-particle interaction.
For realistic systems, which normally do not obey such a limit
one can truncate the equations, which turned out to give very
good results for the Anderson impurity model \cite{And} 
and the spin-Boson model \cite{spBos}. Another approach is to perform
a perturbation expansion in some coupling. This has been applied
to the elimination of the electron-phonon coupling \cite{LeWe}.
Similarly to \cite{GuWe} we will use this approach for the positronium
on the light-cone here.

Before we enter this calculation it seems appropriate to explain,
how this procedure works and to compare it to the similarity renormalization
by Glazek and Wilson \cite{GlWi}. Suppose we would know approximately the
eigenstates of the sector Hamiltonians $PH(l)P$ and $QH(l)Q$ 
and their eigenvalues $E_p(l)$ and $E_q(l)$.
The indices $p$ and $q$ run over all states in the 
$P$- and $Q$-space, respectively. Suppose further, that this basis is 
$l$-independent. This means in other words, that we assume, the off-diagonal
matrix elements $h_{pp'}$ and $h_{qq'}$ of $PHP$ and $QHQ$ are supposed
to be small. Then in evaluating the commutators in Eqs. (\ref{eq:i9})
and (\ref{eq:i10}) we neglect the small off-diagonal matrix elements
$h_{pp'}$ and $h_{qq'}$ and take into account only the diagonal
matrix elements $E_p$ and $E_q$. In Eq. (\ref{eq:i8}) however we
keep all matrix elements on the right-hand side.
Then Eqs.(\ref{eq:i8})-(\ref{eq:i10}) yield
\begin{eqnarray}
   \frac{d h_{pp'}}{dl} &=& 
   \sum_q \Big( \eta_{pq}h_{qp'}-h_{pq}\eta_{qp'} \Big)
\,,\\
   \frac{dh_{pq}}{dl} &=& -\Big( E_p-E_q \Big) \eta_{pq}
\,,\\
   \eta_{pq} &=& \phantom{-}\Big( E_p-E_q \Big) h_{pq} 
\,.\end{eqnarray}
Substituting $\eta_{pq}$ yields
\begin{equation}
   \frac{d h_{pp'}(l)}{dl} = 
   \!\!-\sum_q\!\left(
   \frac{dh_{pq}(l)}{dl}
   \frac{1} {E_p(l)-E_q(l)}h_{qp'}(l)
   +h_{pq}(l)\frac{1} {E_{p'}(l)-E_{q}(l)}
   \frac{dh_{qp'}(l)}{dl} \right)
\,,\label{eq:i16}\end{equation}
where
\begin{eqnarray}
   \frac{dh_{pq}(l)}{dl} &=& -\Big( E_p(l)-E_q(l) \Big)^2 h_{pq}(l)
\,.\label{eq:i17}\end{eqnarray}
The analogous equation for $h_{qq'}$ is obtained 
by interchanging $p\rightarrow q$, $p'\rightarrow q'$.

For the off-diagonal rest part one gets
\begin{equation}
   h_{pq}(l) = h_{pq}(0)\ {\rm exp}
   \Big(-\int_0^l dl'(E_p(l')-E_q(l'))^2 \Big)
\,.\label{eq:i19}\end{equation}
The $l$-dependence in this equation can become important. If in the limit
$l\rightarrow\infty$ the difference $E_p-E_q$ vanishes, then the 
$l$-dependence can be quite crucial. If was first observed in the
spin-Boson model \cite{spBos} and later also for the 
electron-phonon coupling \cite{LeWe}, that a self-consistent solution yields
a decay $E_p(l)-E_q(l)\propto 1/\sqrt{l}$ in the case of asymptotic
degeneracy, so that the corresponding off-diagonal matrix element $h_{pq}$
decays algebraically to zero. This procedure, however, goes beyond perturbation
theory.

One has thus reached the goal:
As the flow parameter tends to infinity the rest sector $PHQ$ 
tends to zero and is eliminated. Simultaneously, this elimination
gives rise to an effective Hamiltonian which has the block-diagonal 
structure $PH(\infty)P$ and $QH(\infty)Q$.
The block $PH(\infty)P$ is defined by
\begin{equation}
   h_{pp'}(\infty) = h_{pp'}(0) -
   \int_0^{\infty}\!\!dl \sum_{q}\left( 
   \frac{dh_{pq}(l)}{dl}\frac{h_{qp'}(l)} {E_p(l)-E_q(l)}
   +  
   \frac{h_{pq}(l)} {E_{p'}(l)-E_{q}(l)}\frac{dh_{qp'}(l)}{dl} 
   \right)
.\label{eq:i18}\end{equation}
The first term represents the initial interaction in $P$-space 
and the second term
originates from the elimination of the off-diagonal rest sectors.

Here we have given a rough idea on how the flow equations work. We have to
consider how they work in perturbation theory and we have to decide on the
blocks defined by the projectors $P$ and $Q$. Obviously the $P$-space should
contain the states with one electron, one positron, and zero photons.
The rest may be covered by the $Q$-space. Since the explicit calculations
are done in terms of creation and annihilation operators, it is easier,
to introduce not two but infinitely many blocks. Each block contains
all states with
a fixed number of electrons, a fixed number of positrons, and a fixed
number of photons. The above equations can be easily generalized to this
case. It is not necessary to write down the blocks explicitly,
since the expressions in terms of creation and annihilation operators
show explicitly, whether the number of particles is conserved or not
and thus, which terms contribute to the diagonal and which to the off-diagonal
part of the Hamiltonian. In quantum electrodynamics the
small coupling is the charge $g$. The leading contribution is 
the kinetic energy of order $g^0$. (The interactions are explicitly
given in the next section). It yields the leading diagonal matrix
elements $E_p$, $E_q$ and no off-diagonal contributions. 
The vertex interaction, which describes emission and absorption of
photons is of order $g$ and purely off-diagonal, since it changes the number
of photons. The instantaneous interaction is of order $g^2$ and contains
both particle-number conserving and particle-number violating contributions.
In zeroth order in $g$ we have only the kinetic energy. This zeroth order
contribution does not change. To first order in $g$ we have 
Eqs. (\ref{eq:i17},\ref{eq:i19}), where the $l$-dependence of $E$ has to be
neglected. Then one enters Eqs. (\ref{eq:i16},\ref{eq:i18}), neglects
the $l$-dependence of $E$ on the right hand-side and obtains $h_{pp'}$
to second order in $g^2$. The result is the effective interaction 
between electrons and positrons. It also includes a change of the 
one-particle energies, which become $l$-dependent in this order.

Equation (\ref{eq:i19}) may be written
\begin{equation}
   h_{pq}(l) = h_{pq}(0)\ f(z_{pq}) 
\,.\end{equation}
with
\begin{equation}
   f(z)=\exp(-z), \quad z_{pq}(l)=\int_0^l dl' (E_p(l')-E_q(l'))^2.
\end{equation}
We discuss now the choice of a more general `similarity function' $f$.
Such a more general function was first by Glazek and Wilson \cite{GlWi}. 

Correspondingly, the generator of the transformation is written as
\begin{equation}
   \eta_{pq}(l) = - \frac{h_{pq}(l)}{E_p(l)-E_q(l)}\frac{d}{dl}
   \Big({\rm ln}f(z_{pq})\Big) 
\,.\label{eq:general}\end{equation}
An example of a different choice for $f$ is given by
\begin{equation}
   \eta_{pq}(l) = \mbox{ sign}(E_p-E_q) h_{pq},
\end{equation}
which yields an exponential decay of $f$
\begin{equation}
   f=\exp(-\int_0^l dl' |E_p(l')-E_q(l')|).
\end{equation}
This similarity function is good, if the sign of 
the difference $E_p-E_q$ does
not depend on the momenta of the interacting particles. This is the
case for the absorption and emission of the photons in the light-cone
frame and can thus be used here. Other possibilities are a sharp cut-off,
if the energy difference is larger than a given energy of $l$. Glazek
and Wilson used a continuous elimination if the energy difference
lies between two energies which decrease with $l$.

Such similarity functions can be used in two cases: \\
(i) In the first case one does not introduce blocks, but aims to
diagonalize single states. Then starting from plane waves the
two-particle interaction becomes negligible, which may prevent the 
procedure from diagonalization. This happened in a first attempt in 
\cite{weg94},
and Jones, Perry, Glazek \cite{JoPeGl} could perform the elimination
of the off-diagonal interaction only down to energy differences of order
Rydberg. This problem which shows up in the continuum, might be
overcome, if one can introduce a discretization. \\
(ii) For block-diagonalization they can be used to the order of perturbation
theory as discussed here. If one goes beyond this order, then it is not
obvious how to use a general similarity function. 
Despite the fact, that there is a lot of
freedom to choose $\eta$, one has to make sure, that the off-diagonal
matrix elements really decay, as shown in Eq. (\ref{eq:i11}) for the choice
(\ref{eq:i5}).

The most important properties of the similarity function $f(z)$ are
\begin{eqnarray}
   f(0) &=& 1
\,,\nonumber\\
   f(z\rightarrow\infty) &=& 0 
\,.\label{eq:ii23}\end{eqnarray}
Its functional dependence on $z$ is less important.
As to be shown below in Sec.~\ref{sec:5} and in Sec.~\ref{sec:6}  
by way of example,
different choices for the similarity function have almost no impact
on physical observables like the spectrum or the wave functions
of bound states.

This leaves us finally with 
\begin{eqnarray}
&& h_{\mbox{eff},pp'} = h_{pp'}(\infty) = h_{pp'}(0)
-\sum_q h_{pq}(0)h_{qp'}(0) \nonumber \\
&& \times \left(
\frac 1{E_p-E_q}\int_0^{\infty}dl\frac{df(z_{pq}(l))}{dl}f(z_{qp'}(l))
+\frac 1{E_{p'}-E_q}\int_0^{\infty}dlf(z_{pq}(l))\frac{df(z_{qp'}(l))}{dl}
\right)
. \label{eq:ii24}\end{eqnarray}
in order $g^2$.

For $p=p'$ this equation may contain ultra-violet divergences
as $l$ goes to 0.
Remember that the elimination of the rest sector in Eq.(\ref{eq:i19}) 
by means of flow equations is performed not in one step,
as in the Tamm-Dancoff approach \cite{tam45,dan50}, 
but rather sequentially for different energy differences, {\it i.e.} 
\begin{eqnarray}
   \lambda=\quad\frac{1} {\sqrt{l}}\leq 
   |E_p-E_q|\leq\frac{1} {\sqrt{l_0\rightarrow 0}}\quad=
   \Lambda\rightarrow\infty
\,.\end{eqnarray}
As it turns out the parameter $\lambda=1/\sqrt{l}$ plays the 
role of an ultra-violet cut-off $\Lambda$ \cite{GlWi}.
The elimination of the matrix element $h_{pq}$ in Eq.(\ref{eq:i19}) 
reminds us to the standard concept of renormalization by Wilson, 
where the high energy modes
are integrated out in the path integral representation
resulting in the effective action for the low energy scales.
Indeed, performing the $l$ ($\lambda$)-integration to the leading order 
one gets for the diagonal elements $h_{pp'}$,
(for $p=p'$ in Eq.(\ref{eq:i16})),
\begin{eqnarray}
   E_p(\lambda) &=& E_p(\Lambda) +
   \sum_q \frac{h_{pq}(\lambda')h_{qp}(\lambda')} 
   {E_p(\Lambda)-E_q(\Lambda)} \ \Bigg\vert^{\lambda}_{\Lambda}
\nonumber\\
   &=& E_p(\Lambda)+\delta E_p(\lambda)-\delta E_p(\Lambda)  
\,.\label{eq:i22}\end{eqnarray}
that defines the connection of energies at different
energy scales and coincides with the second order of conventional 
perturbation theory.
In the case of QED$_{3+1}$, 
as the bare cut-off $\Lambda$ tends to infinity,
the sum in Eq.(\ref{eq:i22}) diverges, 
and one has to introduce the corresponding counter term.
Note that the sum at the upper limit $\lambda$ is regulated
by the similarity factor in $h_{pq}(\lambda)$, since only
the energy differences $\vert E_p-E_q \vert\leq \lambda$ are present.
The ultra-violet renormalization can be attacked with the technique 
of flow equations order by order in a systematic way, 
and further work is in preparation.

In the remainder of this paper, the above schematic equations
of Hamiltonian flow are worked out explicitly for QED$_{3+1}$ 
on the light-cone.

\section{Canonical QED Hamiltonian on the light-front}
\label{sec:3}
Canonical QED$_{3+1}$ in the front form has been 
reviewed recently \cite{bpp97}. Therefore, only 
the most salient features are recollected in this section,
mostly for the purpose to shape notation.
The Lagrangian density for QED 
\begin{eqnarray}
   {\cal L}&=&-\frac{1}{4}F_{\mu \nu} F^{\mu \nu} +
   \overline{\psi}(i \not\! {\partial } +e \not\!\! {A }-
    m )\psi
\end{eqnarray}
is considered here in the light-cone gauge $A^{+}=A^0+A^3=0$. 
Zero modes will be disregarded.
The constrained degrees of freedom, $A^-$ and $\psi_-$
($\Lambda_{\pm}=\frac{1}{2}\gamma^0\gamma^{\pm}$ 
are projection operators, thus $\psi_{\pm}=\Lambda_{\pm}\psi$,
and $\psi=\psi_++\psi_-$)
are removed explicitly and produce the canonical QED Hamiltonian. 
It is  defined through the independent
physical fields $A_{\perp}$ and $\psi_+$ \cite{LeBr}.
To solve the constrained equations for $A^-$ and $\psi_-$  
the auxiliary fields 
\begin{eqnarray}
&& \widetilde A_+ = A_+ - {g\over (i\partial^+)^2}\,J^+ 
\,,\nonumber\\
&& \widetilde\Psi = \Psi _+
   + \left(  m\beta - i\alpha ^i \partial  _{\!\perp i}\right) 
   {1\over 2i\partial _-} \Psi _+
\,,\end{eqnarray}
are introduced. The fermion current is 
$\widetilde J ^{\mu} (x) 
   =\overline{\widetilde\Psi}\gamma^{\mu} \widetilde\Psi$.  
The resulting canonical Hamiltonian $H=P_+$ is given  
as a sum of the free Hamiltonian and the interaction
\begin{equation}
 H=P_+=H_0 + V + W
.\label{I.3} \end{equation}
The free Hamiltonian $H_0$, the `kinetic energy', is 
\begin{equation}
   H_0 = {1\over2}\int\!dx_+d^2x_{\!\perp} 
   \biggl(\overline{\widetilde\Psi} \gamma^+
   {m^2 +(i\nabla_{\!\!\perp}) ^2 \over i\partial^+}
   \widetilde\Psi   +
   \widetilde A ^\mu (i\nabla_{\!\!\perp}) ^2 
   \widetilde A _\mu     \biggr)
\,.\end{equation}
In the `interaction energy' $V + W$, the vertex interaction $V$ is 
the light-cone analogue of  the  minimal coupling interaction
in covariant QED  and $W=W_1 + W_2$ is the sum of the 
instantaneous-gluon $W_1$ and the instantaneous-fermion 
interactions $W_2$.
The latter arise from the constraint equations. 
More explicitly, the interaction is given by
\begin{eqnarray}
   V &=& g  \int\!dx_+d^2x_{\!\perp} 
    \ \widetilde J ^\mu \widetilde A _\mu 
,\nonumber\\
   W_1 &=& { g  ^2 \over 2} \int\!dx_+d^2x_{\!\perp} 
      \ \widetilde J ^+
      {1\over \left(i \partial ^+ \right)^2} \widetilde J ^+ 
,\nonumber\\
   W_2 &=& { g  ^2 \over2} \int\!dx_+d^2x_{\!\perp} 
    \ \overline{\widetilde\Psi}  \gamma ^\mu 
    \widetilde A _\mu \ {\gamma ^+\over i\partial ^+}
    \left( \gamma ^\nu  \widetilde A _\nu 
    \widetilde\Psi  \right)
.\label{I.6}\end{eqnarray}
By definition, the fields 
$\widetilde\Psi=\widetilde\Psi_++\widetilde\Psi_-$ and 
$\widetilde A^\mu = \left( 0,\vec A_{\!\perp },\widetilde A^+ \right)$ 
are the free solutions which in momentum space are parametrized as 
\begin{eqnarray}
   \widetilde\Psi  _{\alpha } (x) &=&
   \sum _\lambda\!\int\!\!  
   {dp^+ d^2p_{\!\bot}\over \sqrt{2p^+(2\pi)^3}}
   \left( b (p) u_\alpha (p ,\lambda ) e^{-ipx} +
   d^\dagger (p)v_\alpha(p,\lambda) 
   e^{+ipx}\right)
,\nonumber \\
   \widetilde A _\mu (x) &=& 
   \sum _\lambda\!\int\!\!  
   {dp^+ d^2p_{\!\bot}\over \sqrt{2p^+(2\pi)^3}}
   \left( a(p)\epsilon_\mu(p,\lambda) e^{-ipx} +
   a^\dagger(p)\epsilon_\mu^\star(p,\lambda)
   e^{+ipx} \right) 
.\end{eqnarray} 
The Dirac spinors and the polarization vectors are
given explicitly in \cite{bpp97}. The single particle 
operators obey the commutation relations 
\begin{equation} 
   \left [ a(p), a^\dagger(p^\prime)\right]  = 
   \left\{ b(p), b^\dagger(p^\prime)\right\} = 
   \left\{ d(p), d^\dagger(p^\prime)\right\} = 
   \delta (p^+-p^{+\,\prime}) 
   \delta ^{(2)}(\vec p_{\!\bot}-\vec p_{\!\bot} ^{\,\prime}) 
   \delta _\lambda ^{\lambda ^\prime}  
\,.\label{I.10}\end{equation}
Inserting the free fields into the Hamiltonian yields 
for the vertex interaction
\begin{eqnarray}
   V &=& g  \int\!dx_+d^2x_{\!\perp}\left.
   \ \overline{\widetilde\Psi}(x)  \gamma ^\mu  
   \widetilde\Psi(x)  \widetilde A _\mu (x) 
   \right\vert_{x^+=0}
\nonumber\\
   &=& {g\over\sqrt{(2\pi)^3}}
   \sum _{\lambda_1,\lambda_2,\lambda_3}
   \int\! {dp^+_1 d^2p_{\!\bot 1}\over \sqrt{2p^+ _1}}
   \int\! {dp^+_2 d^2p_{\!\bot 2}\over \sqrt{2p^+ _2}}
   \int\! {dp^+_3 d^2p_{\!\bot 3}\over \sqrt{2p^+ _3}}
\nonumber\\
   &\times& \int\!{dx_+d^2x_{\!\perp} \over(2\pi)^3} 
   \left[\left( b ^\dagger (p_1) \overline u_\alpha 
   (p_1,\lambda_1) e^{+ip_1x}  +    
    d(p_1)\overline v_\alpha(p_1,\lambda_1)  
   e^{-ip_1x}\right) \right.
\nonumber\\
   &\times&
   \phantom{d^\dagger q}\gamma ^\mu _{\alpha\beta}
   \left.\left( 
    d^\dagger (p_2)v_\beta(p_2,\lambda_2) 
   e^{+ip_2x} +
    b (p_2) u_\beta (p_2 ,\lambda _2) 
   e^{-ip_2x} \right)\right]
\nonumber\\
   &\times& \phantom{\gamma ^\mu _{\alpha\beta}
   d^\dagger q}
   \left.\left( a^\dagger(p_3)\epsilon_\mu^\star
   (p_3,\lambda_3) e^{+ip_3x} +
    a(p_3)\epsilon_\mu(p_3,\lambda_3) 
   e^{-ip_3x} \right) \right.
\,.\label{eq:14}\end{eqnarray} 
The integration over configuration space yields the vertex
interaction as a Fock-space operator. The integration produces
Dirac-delta functions in the single-particle spatial momenta,
\begin{eqnarray}
   \int\!{dx_+\over2\pi} \ e^{i x_+\big(\sum_j p_j^+\big)} 
   &=& \delta\Big( \sum_j p _j^+\Big)
\,,\nonumber\\
   \int\!{d^2x_{\!\perp} \over(2\pi)^2}    
   \ e^{-i\vec x_{\!\perp} \big(\sum_j \vec p_{\!\perp j}\big)}  
   &=& \delta^{(2)}\Big(  \sum_j\vec p_{\!\perp j}\Big)
\,,\end{eqnarray}
and often in the sequel these will be used as a 3-dimensional 
$\delta$-function
\begin{eqnarray}
   \delta^{(3)}\Big(\sum_j p_j\Big)=\delta\Big(\sum_j p_j^+\Big)
 \ \delta^{(2)}\Big(\sum_j \vec{p}_{\perp j}\Big)
\,.\end{eqnarray}
The sums over $j$ run over all respective single particles.
The Dirac-delta's reflect three-momentum conservation,
as always in a Hamiltonian approach.
$W_1$ and $W_2$ as Fock-space operators are obtained correspondingly
and found explicitly in \cite{bpp97}.

\section{Flow equations applied to QED}
\label{sec:4}
In the sequel we consider the canonical Hamiltonian for QED 
as given in Eq.(\ref{I.3})
and work out the details when straightforwardly applying the
flow equations as given in Eqs.(\ref{eq:i8})-(\ref{eq:i10}).
Since we restrict ourselves to solve them  
up to second order in the coupling constant, we can content
ourselves to include explicitly only two Fock-space sectors: 
The sector with one electron and one positron ($e\bar e$)
can be identified with the $P$-space discussed above, 
and the sector with one electron, one positron, 
and one photon ($e\bar e\gamma$) with the $Q$-space.
As a result one aims at the effective Hamiltonian in the
$e\bar e$ space. 
It is helpful to know that the effective interaction must turn 
as the Coulomb potential, to lowest order of approximation.

As a technical trick and for gaining more transparency 
we omit first the instantaneous interactions $W$. 
They will be re-installed at the end of the calculation.
The physical argument is that the instantaneous interaction 
is already of order $g^2$; the changes due to the flow
are of higher order in the coupling constant and have to
be omitted here by consistency.
The light-cone Hamiltonian Eq.(\ref{I.3}) is then
$H=H_0 + V$ and has a very simple structure.
Since the vertex interaction can not have diagonal 
matrix elements, both diagonal sector Hamiltonians 
$PHP$ and $QHQ$ are diagonal operators from the outset.
The case studied here is thus a realization of the
paradigmatic case discussed in the second part of Sec.~\ref{sec:2}.
To the leading order in the coupling 
constant the particle number changing part $PHQ$ is given 
by the vertex interaction, {\it i.e.} $H_r(l)= \widehat V(l)$.
By reasons to become clear soon, we shall put  
hats on the operators in this section. 

At finite flow parameter $l$ the Hamiltonian is 
$H(l)=H_d(l)+H_r(l)$. 
The unitary transformation of the flow equations 
diminishes $\widehat V(l)$ and 
generates a new interaction, $\widehat U(l)$. 
This interaction is (up to second order)
diagonal in particle number and contributes to $H_d(l)=H_0(l)+U(l)$.
The flow equations Eqs.(\ref{eq:i8})-(\ref{eq:i10})
become then consecutively
\begin{eqnarray}
   \frac{d\widehat U(l)}{dl} &=& 
   [\widehat \eta(l),\widehat V(l)]
\,,\label{eq:ii42}\\
   \frac{d\widehat V(l)}{dl} &=& 
   [\widehat \eta(l),\widehat H_0(l)]
\,,\label{eq:ii43}\\
   \widehat \eta (l) &=& [\widehat H_0(l),\widehat V(l)]
\,.\label{eq:ii44}\end{eqnarray}
In this section all of these operators will be evaluated
explicitly.
The free part is 
\begin{eqnarray}
   \widehat H_0(l) = \sum_{\lambda}\int dp^+ d^2p_{\perp} 
   \ E(p;l)
   \left(b^\dagger(p,\lambda)b(p,\lambda)
   +d^\dagger(p,\lambda)d(p,\lambda)
   +a^\dagger(p,\lambda)a(p,\lambda)\right)
.\end{eqnarray}
The single particle energies  ($E=p^-$) depend on the
3-momentum $p=(p^+,\vec p_\perp)$
\begin{equation}
   E(p;l) = \frac{m^2(p;l)+\vec{p}_{\perp}^{\ 2}} {p^+}
\,,\label{eq:21}\end{equation}
and potentially on the flow parameter through
the mass $m^2(p;l)$ of the particle in question.
The interaction term is obtained from Eq.(\ref{eq:14}) 
\begin{eqnarray}
   \widehat V(l) = {1\over\sqrt{(2\pi)^3}} 
   \int\!{[d^3p_1]\over\sqrt{2p^+_1}} 
   \int\!{[d^3p_2]\over\sqrt{2p^+_2}} 
   \int\!{[d^3p_3]\over\sqrt{2p^+_2}}
   \ \delta^{(3)}(p_{1} - p_{2} - p_{2}) \,g(p_1,p_2,p_3;l)
\nonumber\\
   \times  
   \left[ b^\dagger_1 b_2 a_3 \, 
   (\overline u_1 \slash\!\!\!\epsilon_3 u_2) 
   -d_1^\dagger d _2 a_3 \, 
   (\overline v_2 \slash\!\!\!\epsilon_3 v_1)
   +a_1^\dagger d_2 b _3 \, 
   (\overline v_2 \slash\!\!\!\epsilon_1^\star u_3) 
   \right] + h.c.
\,.\label{eq:25}\end{eqnarray}
The integration symbols denote
\begin{equation}
   [d^3p]= dp^+ d^2p_{\perp}\sum _{\lambda}
\,,\end{equation}
for example, and the abbreviations 
$u_1\equiv u(p_1,\lambda_1)$ and 
$\slash\!\!\!\epsilon_3^\star
 \equiv \gamma^\mu\epsilon_\mu^\star(p_3,\lambda_3)$
are introduced for the sake of a compact notation. 
The effective coupling `constant' has the initial value
\begin{equation}
   g(p_1,p_2,p_3;l=0)= g = e
\,,\label{eq:ii49}\end{equation}
with the fine structure constant $\alpha=e^2/4\pi\sim 1/137$.
Correspondingly, the generator of the unitary transformation 
$\widehat \eta = [\widehat H_0,\widehat V]$ becomes 
\begin{eqnarray}
   \widehat \eta (l) = {1\over\sqrt{(2\pi)^3}}
   \int\!{[d^3p_1]\over\sqrt{2p^+_1}} 
   \int\!{[d^3p_2]\over\sqrt{2p^+_2}} 
   \int\!{[d^3p_3]\over\sqrt{2p^+_3}}
   \,\delta^{(3)}(p_{1} - p_{2} - p_{3}) \, \eta (p_1,p_2,p_3;l)
\nonumber\\
   \times  
   \left[ b^\dagger_1 b_2 a_3 \, 
   (\overline u_1 \slash\!\!\!\epsilon_3 u_2) 
   -d_1^\dagger d _2 a_3 \, 
   (\overline v_2 \slash\!\!\!\epsilon_3 v_1)
   +a_1^\dagger d_2 b _3 \, 
   (\overline v_2 \slash\!\!\!\epsilon_1^\star u_3) 
   \right] - h.c.
\,.\label{eq:28}\end{eqnarray}
The structure of $\widehat \eta$ is very similar
to $\widehat V$ because $\widehat H_0$ is diagonal, thus
\begin{equation}
   \eta (p_1,p_2,p_3;l) = g(p_1,p_2,p_3;l)D(p_1,p_2,p_3;l)
\,.\end{equation}
It is convenient to introduce the difference of single particle energies 
\begin{equation}
   D(p_1,p_2,p_3;l) = E(p_1;l) - E(p_2;l) - E(p_3;l)
\,,\end{equation}
with the $E$'s being defined in Eq.(\ref{eq:21}).
The derivative 
$d\widehat V (l)/dl = [\widehat \eta(l),\widehat H _0(l)]$
becomes
\begin{eqnarray}
   \frac{d\widehat V (l)}{dl}  = - 
   \int\!{[d^3p_1]\over\sqrt{2p^+_1}} 
   \int\!{[d^3p_2]\over\sqrt{2p^+_2}}  
   \int\!{[d^3p_3]\over\sqrt{2p^+_2}} 
   \,\delta^{(3)}(p_{1} - p_{2} - p_{3}) 
   \,{\eta(p_1,p_2,p_3;l)\over\sqrt{(2\pi)^3}}
\nonumber\\
   \times D(p_1,p_2,p_3;l) \left[ b^\dagger_1 b_2 a_3 \, 
   (\overline u_1 \slash\!\!\!\epsilon_3 u_2) 
   -d_1^\dagger d _2 a_3 \, 
   (\overline v_2 \slash\!\!\!\epsilon_3 v_1)
   +a_1^\dagger d_2 b _3 \, 
   (\overline v_2 \slash\!\!\!\epsilon_1^\star u_3) 
   \right] 
   + h.c.
\,.\end{eqnarray}
Finally, we calculate the new interactions
$\widehat U(l)$ which are defined through the derivative 
${d\widehat U(l)}/{dl} = [\widehat \eta(l), \widehat V(l)]$.
Their calculation is somewhat cumbersome but straightforward. 
Inserting the six terms from Eq.(\ref{eq:25}) and the six terms
from Eq.(\ref{eq:28}) gives 36 terms for $\widehat U(l)$ 
which by symmetries reduce to six interactions 
in the 2-particle sectors.
In the present work one restricts to calculate 
$\widehat  U_{e\bar e}$, 
the effective interaction between an electron and a positron,
as illustrated in Fig.~\ref{fig:1}.
It has an exchange part and an annihilation part. 

\subsection{The exchange part}

\input fig12.tex
The matrix element $\widetilde U_{ex}$ of the exchange part 
\begin{eqnarray}
   \widehat  U_{e\bar e;ex}(l) &=& 
   \int\![d^3p_1] \int\! [d^3p_2] \int\! [d^3p'_1] \int\! [d^3p'_2]
   \ \delta^{(3)}(p_1+p_2-p'_1-p'_2)
\nonumber\\ && 
   \widetilde U_{ex}(p_1,p_2;p'_1,p'_2;l)
   \ b^\dagger (p_1,\lambda_1) b(p'_1,\lambda'_1)
   \,d^\dagger (p_2,\lambda_2) d(p'_2,\lambda'_2) 
\,.\label{eq:33}\end{eqnarray}
is calculated next to some detail.
By direct substitution of Eqs.(\ref{eq:25}) and (\ref{eq:28})
one gets 
\begin{eqnarray}
   \frac{d\widehat  U_{e\bar e;ex}^{gen}(l)}{dl} = -{1\over(2\pi)^3} 
   \int\!{[d^3p_1]\over\sqrt{2p^+_1}}
   \int\!{[d^3p_2]\over\sqrt{2p^+_2}} 
   \int\!{[d^3p_3]\over\sqrt{2p^+_3}}
   \int\!{[d^3p_1^\prime]\over\sqrt{2p^{\prime+}_1}} 
   \int\!{[d^3p_2^\prime]\over\sqrt{2p^{\prime+}_2}} 
   \int\!{[d^3p_3^\prime]\over\sqrt{2p^{\prime+}_3}}
\nonumber\\ 
   \Big(
   \,\delta^{(3)}(p_1 - p'_1 - p_3) 
   \,\delta^{(3)}(p'_2 - p_2 - p'_3) 
   \,\left[b_1^\dagger b_{1'} a_3,
   a_{3'}^\dagger d_{2}^\dagger d _{2'}\right]
   (\overline u_1\slash\!\!\!\epsilon _3 u _{1'})\,
   (\overline v_{2'}
   \slash\!\!\!\epsilon^\star_{3'} v_{2}) 
\nonumber\\ 
   \big(\eta(p_1,p'_1,p_3;l)\,g(p'_2,p_2,p'_3;l) + 
   \eta(p'_2,p_2,p'_3;l)\,g(p_1,p'_1,p_3;l) \big)
\nonumber\\ 
   +
   \,\delta^{(3)}(p'_1 - p_1 - p_3) 
   \,\delta^{(3)}(p_2 - p'_2 - p'_3) 
   \left[d_{2}^\dagger d _{2'} a_{3'} ,
   a_{3}^\dagger b_{1}^\dagger b_{1'} \right]
   (\overline u_{1}\slash\!\!\!\epsilon_{3}^\star u_{1'}) 
   (\overline v_{2'}\slash\!\!\!\epsilon_{3'} v_{2}) \,
\nonumber\\  
   \big( \eta(p'_1,p_1,p_3;l)\,g(p_2,p'_2,p'_3;l)+ 
   \eta(p_2,p'_2,p'_3;l)\,g(p'_1,p_1,p_3;l)\big) \Big)
\,.\label{eq:32}\end{eqnarray}
The commutation relations Eq.(\ref{I.10}) for the photon induce 
a three-momentum delta-function
$\delta^{(3)}(p_{3} - p_{3}^\prime)$. 
Since
\begin{eqnarray}
   \eta(p_1,p'_1,p_1-p'_1;l) &=& - \eta(p'_1,p_1,p'_1-p_1;l)
\,,\nonumber\\ 
   g(p'_1,p_1,p'_1-p_1;l)  &=& \phantom{-}g(p_1,p'_1,p_1-p'_1;l) 
\,,\end{eqnarray} 
all of the integrations in Eq.(\ref{eq:32}) can be performed trivially.
The sum over the photon helicity is carried out by introducing 
the polarization tensor \cite{LeBr,bpp97}
\begin{eqnarray}
   d_{\mu\nu}(q)\equiv \sum_{\lambda}
   \epsilon_{\mu}(q,\lambda)\epsilon^{*}_{\nu}(q,\lambda) =
   - g_{\mu\nu} + 
   \frac{\eta_{\mu}q_{\nu}+\eta_{\nu}q_{\mu}}{q^+}
\,.\label{eq:32a}\end{eqnarray}
The null vector $\eta^{\mu}$ has the components
$(\eta^+,\vec\eta_\perp, \eta^-)=(0,\vec{0},2)$
and should not be confused with the generator $\eta$.  
Dropping hence forward the argument $l$ in $g$ or $\eta$
for the reason of notational compactness 
the $l$-derivative of $\widetilde U_{ex}^{gen}$ becomes 
\begin{eqnarray}
   {d\widetilde U_{ex}^{gen} \over dl} && = - {1\over 2(2\pi)^3}
   \left( \theta({p_1}^+ - {p'_1}^+) + \theta({p'_1}^+ - {p_1}^+) 
   \right)
\nonumber\\ &\times& 
   {(\overline u(p_1,\lambda_1)\gamma^\mu u(p'_1,\lambda'_1))
   \over\sqrt{2p_1^{+}}\ \sqrt{2p_1^{\prime+}} }
   {d_{\mu\nu}(q)\over  q^+ }   
   {(\overline v(p'_2,\lambda'_2)\gamma^{\nu} v(p_2,\lambda_2)) 
   \over\sqrt{2p_2^{\prime+}}\ \sqrt{2p_2^{+}}}
\nonumber\\ &\times& 
   \left(\eta(p'_1,p_1,p'_1-p_1) g(p_2,p'_2,p_2-p'_2)  + 
    g(p'_1,p_1,p'_1-p_1) \eta(p_2,p'_2,p_2-p'_2) \right)
.\label{eq:ii58}\end{eqnarray} 
The 4-momentum of the photon is denoted by $q^\mu$, see also
Eq.(\ref{eq:ii66}) below. 
The step function $\theta(x)=1$ for $x\ge 0$, 
and zero otherwise. Since $\theta(x)+\theta(-x)=1$
the sum of the two theta-functions will be replaced by unity
in the sequel. This reminds to the calculation 
of the $q\bar q$-scattering amplitude which was given explicitly
in \cite{bpp97}. 
To get the full effective exchange interaction
one has to include also the instantaneous exchange interaction 
\begin{equation}
   \widetilde U_{ex}^{inst}(l) = - \frac{e^2}{2(2\pi)^3}
   \frac{(\overline u(p_1,\lambda_1) \gamma^\mu
       u(p'_1,\lambda'_1))}{\sqrt{2p_1^+}\sqrt{2p_1^{'+}}}
   \frac{(\overline v(p'_2,\lambda'_2)\gamma^\nu
    v(p_2,\lambda_2))}{\sqrt{2p_2^{'+}}\sqrt{2p_2^+}}
   \frac{\eta_\mu \eta_\nu}{q^{+2}}
\,.\label{eq:46}\end{equation}
To the order considered here it is independent of $l$,
{\it i.e.}  $d\widetilde U_{ex}^{inst}/dl=0$. 

\subsection{The annihilation part}

Having been so explicit for the exchange part one can
proceed rather quickly for the annihilation channel
where the calculation proceeds quite correspondingly. 
One defines first the matrix element $\widetilde U_{an}$ by
\begin{eqnarray}
   \widehat  U_{e\bar e;an}(l) &=& 
   \int\![d^3p_1] \int\! [d^3p_2] \int\! [d^3p'_1] \int\! [d^3p'_2]
   \ \delta^{(3)}(p_1+p_2-p'_1-p'_2)
\nonumber\\ && 
   \widetilde U_{an}(p_1,p_2;p'_1,p'_2;l)
   \ b^\dagger (p_1,\lambda_1) b(p'_1,\lambda'_1) 
   \,d^\dagger (p_2,\lambda_2) d(p'_2,\lambda'_2) 
\,.\label{eq:33a}\end{eqnarray}
Its $l$-derivative is defined by the flow equations, {\it i.e.} 
\begin{eqnarray}
   \frac{d\widehat  U_{e\bar e;an}^{gen}(l)}{dl} = -{1\over(2\pi)^3} 
   \int\!{[d^3p_1]\over\sqrt{2p^+_1}}
   \int\!{[d^3p_2]\over\sqrt{2p^+_2}} 
   \int\!{[d^3p_3]\over\sqrt{2p^+_3}}
   \int\!{[d^3p_1^\prime]\over\sqrt{2p^{\prime+}_1}} 
   \int\!{[d^3p_2^\prime]\over\sqrt{2p^{\prime+}_2}} 
   \int\!{[d^3p_3^\prime]\over\sqrt{2p^{\prime+}_3}}
\nonumber\\ 
   \,\delta^{(3)}(p_1 + p_2 - p_3) 
   \,\delta^{(3)}(p'_1 + p'_2 - p'_3) 
   \,\left[b_1^\dagger d_2^\dagger a_3,
   a_{3'}^\dagger d_{2'} b_{1'}\right]
   (\overline u_1\slash\!\!\!\epsilon _3 v _{2})\,
   (\overline v_{2'}
   \slash\!\!\!\epsilon^\star_{3'} u_{1'}) 
\nonumber\\ 
   \Big(\big(\eta(p_3,p_2,p_1)\,g(p'_3,p'_2,p'_1) + 
   \eta(p'_3,p'_2,p'_1)\,g(p_3,p_2,p_1) \big) \Big)
\,.\label{eq:34a}\end{eqnarray}
All integrations can be performed explicitly and one arrives at
\begin{eqnarray}
   {d\widetilde U_{an}^{gen}\over dl} &=& - {1\over 2(2\pi)^3}\,
   {(\overline u(p_1,\lambda_1)\gamma^\mu v(p_2,\lambda_2))
   \over\sqrt{2p_1^{+}}\ \sqrt{2p_2^{+}} }
   {d_{\mu\nu}(p)\over  p^+ }   
   {(\overline v(p'_2,\lambda'_2)\gamma^{\nu} u(p'_1,\lambda'_1)) 
   \over\sqrt{2p_2^{\prime+}}\ \sqrt{2p_1^{\prime+}}}
\nonumber\\ &\times& 
   \left(\eta(p_1+p_2,p_2,p_1) g(p'_1+p'_2,p'_2,p'_1) + 
   g(p_1+p_2,p_2,p_1) \eta(p'_1+p'_2,p'_2,p'_1) \right)
.\end{eqnarray} 
The 4-momentum of the photon is denoted here by $p$.
The instantaneous interaction in the exchange channel
\begin{eqnarray}
   \widetilde U_{an}^{inst}(l) &=& \frac{e^2}{2(2\pi)^3}
   \frac{(\overline u(p_1,\lambda_1)\gamma^\mu
   v(p_2,\lambda_2))}{\sqrt{2p_1^+}\sqrt{2p_2^+}}
   \frac{(\overline v(p'_2,\lambda'_2)\gamma^\nu
   u(p'_1,\lambda'_1))}{\sqrt{2p_2^{'+}}\sqrt{2p_1^{' +}}}
   \frac{\eta_\mu\eta_\nu}{p^{+2}}
\label{eq:54}\end{eqnarray}
is again independent of $l$ in the lowest non-trivial order of the
coupling constant.

\subsection{Integrating the flow equations for the exchange}

The first order flow equations have been given in Eqs.(\ref{eq:ii43}) 
and (\ref{eq:ii44}) in operator form. After evaluating
all matrix elements they reduce simply to two coupled equations: 
\begin{eqnarray}
   \frac{dg(p_1,p_2,p_3;l)}{dl} &=& -
   D(p_1;p_2,p_3;l)\ \eta(p_1,p_2,p_3;l)
\,,\label{eq:37}\\ 
   \eta(p_1,p_2,p_3;l) &=& \phantom{-}
   D(p_1,p_2,p_3;l)\ g(p_1,p_2,p_3;l)
\,.\label{eq:38}\end{eqnarray}
Replacing $g$ by the suitably normalized similarity function $f$
according to 
\begin{equation}
   g(p_1,p_2,p_3;l)=g(0)f(p_1,p_2,p_3;l)=
   e f(p_1,p_2,p_3;l)
\,,\label{eq:40}\end{equation}
see also Eq.(\ref{eq:ii49}),
one gets 
\begin{equation}
   f(p_1,p_2,p_3;l) = 
   {\rm exp} \left(-l\ D^2(p_1,p_2,p_3)\right)
\label{eq:f67}\end{equation}
as the explicit solution. It describes 
the decay rate of the off-diagonal vertex interaction $V$.
However, because of the considerations in Sec.~\ref{sec:2} 
particularly Eq.(\ref{eq:general}) and the considerations below
we want to keep $f=f(D;l)$ as a general function. 
Correspondingly, we rewrite $\eta$ as 
\begin{equation}
   \eta(l) = - \frac{1}D
   \left(\frac{d{\rm ln}f(D;l)}{dl}\right)g(l)
\,.\label{eq:55a}\end{equation}
The formal integration of the flow Eqs. (\ref{eq:37}) and 
(\ref{eq:38}) can be treated
more compactly in a reasonably short notation.
We therefore define the always negative quantities
\begin{eqnarray}
   D_{e} &=& {p'_1}^- - p_1^{-} - (p'_1-p_1)^-
\,,\nonumber\\
   D_{\bar e} &=& p_2^{-} - {p'_2}^- - (p_2-p'_2)^-
\,.\label{eq:40a} \end{eqnarray}
They represent the energy differences along the electron 
and the positron line, respectively, and are in simple relationship 
to both the 4-momentum of the exchanged photon
\begin{equation}
   q_\mu = p'_{1\mu}-p_{1\mu}-\eta_\mu \frac{D_{e}}{2}
         = p_{2\mu}-p'_{2\mu}-\eta_\mu \frac{D_{\bar e}}{2}
\,,\label{eq:ii66}\end{equation}
and to the (Feynman-) 4-momentum transfers along the two lines 
\begin{eqnarray}
   Q_{e}^2 &=& -(p'_1-p_1)^2 = -q^+D_{e} 
\,,\\
   Q_{\bar e}^2 &=& -(p_2-p'_2)^2 = -q^+D_{\bar e}
\,,\end{eqnarray}
which need not be equal in a Hamiltonian approach.
Since the always positive Feynman\--momentum transfer $Q$ 
is a more physical quantity than the energy difference, 
the $D$'s will be substituted in the sequel by the $Q$'s 
as long as no misunderstanding can arise. 
In fact, we shall use often the 
{\em mean-square momentum transfer}
and the {\em mean-square difference}
\begin{eqnarray}
   Q^2 &=& {1\over 2}(Q_{e}^2+Q_{\bar e}^2)
   = -{q^+\over 2}(D_{e}+D_{\bar e})
\,,\\  \delta
   Q^2 &=& {1\over 2}(Q_{e}^2-Q_{\bar e}^2)
   = -{q^+\over 2}(D_{e}-D_{\bar e})
\,,\label{eq:r15}\end{eqnarray}
respectively.
The above definitions are also useful to simplify the
polarization tensor appearing in Eq.(\ref{eq:ii58}).
Since $d_{\mu\nu}(q)$ appears always in combinations with the 
spinors one can make use of the Dirac equation 
$(p_1-p'_1)_\mu \overline u(p_1)\gamma^\mu u(p'_1) = 0$
and write
\begin{eqnarray}
   q_\mu \overline u(p_1,\lambda_1)\gamma^\mu u(p'_1,\lambda'_1) 
   = - {D_{e}\over 2}
   \eta_\mu \,\overline u(p_1,\lambda_1)\gamma^\mu u(p'_1,\lambda'_1)
.\end{eqnarray} 
One can replace thus in Eq.(\ref{eq:ii58})
$d_{\mu\nu}(q)\longrightarrow -g_{\mu\nu}+\eta_\mu \eta_\nu (Q/q^+)^2$.
With these definitions we return now to the problem of
integrating up Eq.(\ref{eq:ii58}). Substituting
Eqs.(\ref{eq:37}) and (\ref{eq:40}) one has 
for its $l$-dependent part
\begin{eqnarray}
   & & \eta(p'_1,p_1,p'_1-p_1;l)g(p_2,p'_2,p_2-p'_2;l)
   +\eta(p_2,p'_2,q;l)g(p'_1,p_1,q;l) 
\nonumber\\
   &=& - e^2 \left(\frac{1}{D_{e}}
   \frac{d f (D_{e};l)}{dl} f (D_{\bar e};l)
   +\frac{1}{D_{\bar e}}\frac{d f (D_{\bar e};l)}{dl} 
   f (D_{e};l)\right)
\,.\label{eq:ii72}\end{eqnarray} 
For the formal integration it turns out useful to introduce
the abbreviation
\begin{equation}
   \Theta(D_{e},D_{\bar e}) = - \int_{0}^\infty\!dl'\,
   \frac{d f (D_{e};l')}{dl'} f (D_{\bar e};l')
   \equiv \Theta_{e\bar e}
\,,\label{eq:ii77}\end{equation}
which is not symmetric in the arguments but which satisfies 
by means of Eq.(\ref{eq:ii23}) 
\begin{equation}
   \Theta(D_{e},D_{\bar e})+\Theta(D_{\bar e},D_{e}) =
   \Theta_{e\bar e} + \Theta_{\bar e e} = 1
\,.\label{eq:ii18a}\end{equation}
When $l$-integrating Eq.(\ref{eq:ii72}), 
the $l$-dependence of $D_i(l)$ in the denominator
can be neglected to the order considered here, 
therefore
\begin{eqnarray}
   &&\int_0^\infty\!dl'\left(
   \eta(p'_1,p_1,p'_1-p_1;l')g(p_2,p'_2,p_2-p'_2;l')
   +\eta(p_2,p'_2,q;l')g(p'_1,p_1,q;l') \right)
\nonumber\\
   &&= e^2 
   \left(\frac{\Theta_{e\bar e}}{D_{e}}
   +\frac{\Theta_{\bar e e}} {D_{\bar e}}\right) 
   = - e^2 q^+ \left(
   \frac {\Theta_{e\bar e}}  {Q_{e}^2} +
   \frac {\Theta_{\bar e e}} {Q_{\bar e}^2}\right) 
.\end{eqnarray} 
The latter combination appears repeatedly, see for example 
Eq.(\ref{eq:ii18b}).
Putting things together the generated  
interaction Eq.(\ref{eq:ii58}) becomes
\begin{eqnarray}
   \widetilde U_{ex}^{gen} = &+&\frac {e^2} {2(2\pi)^3}
   \frac {(\overline u(p_1,\lambda_1) \gamma^\mu
       u(p'_1,\lambda'_1))} {\sqrt{2p_1^+}\sqrt{2{p'_1}^+}}
   \frac {(\overline v(p'_2,\lambda'_2)\gamma^\nu
    v(p_2,\lambda_2))} {\sqrt{2{p'_2}^+}\sqrt{2p_2^+}}
\nonumber\\
   &\times& 
   d_{\mu\nu}(q)
   \left(\frac{\Theta_{e \bar e}} {Q_{e}^2} +
         \frac{\Theta_{\bar e e}} {Q_{\bar e}^2}\right)
\,,\label{eq:ii81}\end{eqnarray}
and after substituting $d_{\mu\nu}(q)$
\begin{eqnarray}
   \widetilde U_{ex}^{gen} = &-&\frac {e^2} {2(2\pi)^3}
   \frac {(\overline u(p_1,\lambda_1) \gamma^\mu
       u(p'_1,\lambda'_1))} {\sqrt{2p_1^+}\sqrt{2{p'_1}^+}}
   \frac {(\overline v(p'_2,\lambda'_2)\gamma^\nu
    v(p_2,\lambda_2))} {\sqrt{2{p'_2}^+}\sqrt{2p_2^+}}
\nonumber\\
   &\times& 
   \left(g_{\mu\nu}-\eta_\mu \eta_\nu\frac {Q^2}{{q^+}^2}\right)
   \left(\frac{\Theta_{e \bar e}} {Q_{e}^2} +
         \frac{\Theta_{\bar e e}} {Q_{\bar e}^2}\right)
\,.\label{eq:r81}\end{eqnarray}
Adding the instantaneous exchange interaction
Eq.(\ref{eq:46}), using Eq.(\ref{eq:ii18a}), 
one gets 
\begin{eqnarray}
   \widetilde U_{ex} = &-& \frac{e^2}{2(2\pi)^3}
   \frac{(\overline u(p_1,\lambda_1) \gamma^\mu
       u(p'_1,\lambda'_1))}{\sqrt{2p_1^+}\sqrt{2{p'_1}^{+}}}
   \frac{(\overline v(p'_2,\lambda'_2)\gamma^\nu
    v(p_2,\lambda_2))}{\sqrt{2{p'_2}^+}\sqrt{2p_2^+}} 
\nonumber\\
   &\times& \left[g_{\mu\nu}
   \left(\frac{\Theta_{e \bar e}} {Q_{e}^2} +
         \frac{\Theta_{\bar e e}} {Q_{\bar e}^2}\right) + 
   \eta_\mu\eta_\nu
   \left(\frac{\Theta_{e \bar e}} {Q_{e}^2} - 
         \frac{\Theta_{\bar e e}} {Q_{\bar e}^2}\right)\,
   \frac {\delta Q^2}{{q^+}^2}
   \right]
\label{eq:47}\end{eqnarray}
for the matrix element of the total exchange interaction.

\subsection{Integrating the flow equations for the annihilation}

For the annihilation term we define the 
(now always positive) energy differences as
\begin{eqnarray}
   D_a &=& {p'_1}^- + {p'_2}^- - (p'_1+p'_2)^-
\nonumber\\
   D_b &=& p_1^{-} + p_2^{-} - (p_1+p_2)^-
\,.\label{eq:40b} \end{eqnarray}
They are related to the 4-momentum of the photon $p^\mu$ 
in the $t$-channel by
\begin{equation}
   p_\mu = p'_{1\mu}+p'_{2\mu}-\eta_\mu \frac{D_a}{2}
         = p _{1\mu}+p _{2\mu}-\eta_\mu \frac{D_b}{2}
\,.\end{equation}
Rather than the momentum transfer $Q$ the free invariant 
mass-squares of the initial and final states 
are introduced by
\begin{eqnarray}
   M_a^2 &=& (p'_1+p'_2)^2 = p^+ D_a 
\,,\\
   M_b^2 &=& (p _1+p _2)^2 = p^+ D_b
\,,\end{eqnarray}
as well as their mean and difference
\begin{eqnarray}
   M^2 &=& {1\over 2}(M_a^2 + M_b^2)= {p^+\over 2}(D_a+D_b)
\,,\\  \delta
   M^2 &=& {1\over 2}(M_a^2 - M_b^2)= {p^+\over 2}(D_a-D_b)
\,,\end{eqnarray}
respectively. The generated annihilation interaction becomes 
\begin{eqnarray}
   \widetilde U_{an}^{gen} &=& +\frac{e^2}{2(2\pi)^3}
   \frac{(\overline u(p_1,\lambda_1)\gamma^\mu
   v(p_2,\lambda_2))}{\sqrt{2p_1^+}\sqrt{2p_2^+}}
   \frac{(\overline v(p'_2,\lambda'_2)\gamma^\nu
   u(p'_1,\lambda'_1))} {\sqrt{2{p'_2}^+}\sqrt{2{p'_1}^+}} 
\nonumber\\
   &\times& d_{\mu\nu}(p) 
   \left(\frac{\Theta_{ab}} {M_a^2} +
         \frac{\Theta_{ba}} {M_b^2} \right) 
\,.\end{eqnarray}
The Dirac equation 
$(p_1+p_2)_\mu \overline u(p_1)\gamma^\mu v(p_2) = 0$
allows to write 
\begin{eqnarray}
   p_\mu \overline u(p_1,\lambda_1)\gamma^\mu v(p_2,\lambda_2) 
   = - {D_b\over 2}
   \eta_\mu \,\overline u(p_1,\lambda_1)\gamma^\mu v(p_2,\lambda_2)
\,.\end{eqnarray} 
After substituting 
$d_{\mu\nu}(p)\longrightarrow 
 -g_{\mu\nu}-\eta_\mu \eta_\nu M^2/{p^+}^2$ 
 one gets for $\widetilde U_{an}^{gen}$
\begin{eqnarray}
   \widetilde U_{an}^{gen} = &-&\frac{e^2}{2(2\pi)^3}
   \frac{(\overline u(p_1,\lambda_1)\gamma^\mu
   v(p_2,\lambda_2))}{\sqrt{2p_1^+}\sqrt{2p_2^+}}
   \frac{(\overline v(p'_2,\lambda'_2)\gamma^\nu
   u(p'_1,\lambda'_1))} {\sqrt{2{p'_2}^+}\sqrt{2{p'_1}^+}} 
\nonumber\\
   &\times&
   \left(g_{\mu\nu} + \eta_\mu \eta_\nu 
   \frac {M^2} {{p^+}^2}\right)
   \left(\frac{\Theta_{ab}} {M_a^2} +
         \frac{\Theta_{ba}} {M_b^2} \right) 
\,.\end{eqnarray}
Adding the instantaneous term yields
the effective interaction in the annihilation channel
\begin{eqnarray}
   \widetilde U_{an} = &-& \frac{e^2}{2(2\pi)^3}
   \frac{(\overline u(p_1,\lambda_1)\gamma^\mu
   v(p_2,\lambda_2))}{\sqrt{2p_1^+}\sqrt{2p_2^+}}
   \frac{(\overline v(p'_2,\lambda'_2)\gamma^\nu
   u(p'_1,\lambda'_1))} {\sqrt{2{p'_2}^+}\sqrt{2{p'_1}^+}} 
\nonumber\\
   &\times& \left[
   g_{\mu\nu}
   \left(\frac{\Theta_{ab}} {M_a^2} +
         \frac{\Theta_{ba}} {M_b^2} \right) - 
   \eta_\mu\eta_\nu
   \left(\frac{\Theta_{ab}} {M_a^2} - 
         \frac{\Theta_{ba}} {M_b^2} \right) \,
         \frac{\delta M^2} {{p^+}^2} \right]
\,,\label{eq:55}\end{eqnarray}
all in perfect analogy to the exchange term.

\subsection{The effective $e\bar e$-interaction}

Thus far, we have been studying the structure
of the Hamiltonian proper $H=P_+=P^-/2$.
In dealing with its spectra it is advantageous to 
study the spectrum of the `light-cone Hamiltonian' 
$H_{\rm LC}=P_\mu P^\mu=P^+P^- - P_{\!\perp}^{\,2}$. 
It is a Lorentz scalar \cite{bpp97}
with the eigenvalues having the dimension of an 
invariant mass-squared.
Combining the effective interaction
of the exchange and annihilation channels
one introduces therefore 
\begin{equation}
   U_{\rm eff} = {P^+}^2 (\widetilde U_{ex} + \widetilde U_{an})
\,,\end{equation}
see Eq.(\ref{eq:47}) and Eq.(\ref{eq:55}). One gets 
\begin{eqnarray}
   U_{\rm eff} = -  
   \frac{\alpha}{4\pi^2} 
   <\gamma^\mu\gamma^\nu>_{ex}
   \left[g_{\mu\nu}\,
   \left(\frac{\Theta_{e \bar e}} {Q_{e}^2} +
         \frac{\Theta_{\bar e e}} {Q_{\bar e}^2}\right) + 
   \eta_\mu\eta_\nu\,\frac {\delta Q^2}{{q^+}^2} 
   \left(\frac{\Theta_{e \bar e}} {Q_{e}^2} - 
         \frac{\Theta_{\bar e e}} {Q_{\bar e}^2}\right)
   \right]\phantom{.}
\nonumber\\
  - \frac{\alpha}{4\pi^2} 
  <\gamma^\mu\gamma^\nu>_{an}
  \left[
   g_{\mu\nu}
   \left(\frac{\Theta_{ab}} {M_a^2} +
         \frac{\Theta_{ba}} {M_b^2} \right) - 
   \eta_\mu\eta_\nu\,\frac{\delta M^2} {{p^+}^2}
   \left(\frac{\Theta_{ab}} {M_a^2} - 
         \frac{\Theta_{ba}} {M_b^2} \right)  \right]
.\label{eq:56}\end{eqnarray} 
The symbols $<\gamma^\mu\gamma^\nu>$ are introduced conveniently as 
\begin{eqnarray}
   <\gamma^\mu\gamma^\nu>_{ex} = 
   \frac{(\overline u(p_1,\lambda_1) \gamma^\mu u(p'_1,\lambda'_1))\,
   (\overline v(p'_2,\lambda'_2) \gamma^\nu v(p_2,\lambda_2))}
   {\sqrt{xx'(1-x)(1-x')}}
   = \frac {\left(j^\mu J^\nu\right)_{ex}} {\sqrt{xx'(1-x)(1-x')}}
,\nonumber\\
   <\gamma^\mu\gamma^\nu>_{an} =
   \frac{(\overline u(p_1,\lambda_1)\gamma^\mu  v(p_2,\lambda_2))\,
   (\overline v(p'_2,\lambda'_2)\gamma^\nu u(p'_1,\lambda'_1))}
   {\sqrt{xx'(1-x)(1-x')}}
   = \frac {\left(c^\mu C^\nu\right)_{an}} {\sqrt{xx'(1-x)(1-x')}}
,\label{eq:r94}\end{eqnarray} 
where for example $x = p_1^+/P^+$ is the 
longitudinal momentum fraction \cite{bpp97}.  

The effective interaction $U_{\rm eff}$ 
is the kernel of the integral equation
\begin{eqnarray} 
    M^2\ \langle x,\vec k_{\!\perp}; \lambda_{1},
    \lambda_{2}  \vert \psi \rangle =
    {m^2 + \vec k_{\!\perp}^2 \over x(1-x)} 
    \ \langle x,\vec k_{\!\perp}; \lambda_{1},
    \lambda_{2}  \vert \psi \rangle \phantom{.}
\nonumber\\ 
    +\sum _{ \lambda_{1}^\prime,\lambda_{2}^\prime}
    \!\int_D\!dx^\prime d^2 \vec k_{\!\perp}^\prime\,
    \,\langle x,\vec k_{\!\perp}; \lambda_{1}, \lambda_{2}
    \vert U_{\rm eff}
    \vert x^\prime,\vec k_{\!\perp}^\prime; 
    \lambda_{1}^\prime, \lambda_{2}^\prime\rangle
    \ \langle x^\prime,\vec k_{\!\perp}^\prime; 
    \lambda_{1}^\prime,\lambda_{2}^\prime  
    \vert \psi \rangle
\,.\label{eq:r96}\end {eqnarray}
In the equation appear only intrinsic transversal momenta 
$\vec k_{\!\perp}$ and longitudinal momentum fractions 
$x = p_1^+/P^+$, defined by
{\it i.e.} $p_1^\mu=(xP^+,x\vec P_\perp + \vec k_{\!\perp},p_1^-)$.
Its spectrum is thus manifestly independent of the kinematical state
of the bounded system,
particularly of $P^+$ and $\vec P_\perp$, which reflects the boost 
invariances peculiar to the front form \cite{bpp97}.
The integral equation replaces in some way Eq.(\ref{eq:i1}). 
The first term on the r.h.s is the free part of the
Hamiltonian in analogy to a `kinetic energy', and the
second term is an  `interaction energy' 
which is the relativistically correct interaction,  
correct up to the second order in the coupling constant.

The integration domain $D$ is restricted by the covariant cut-off
condition of Brodsky and Lepage \cite{LeBr},
\begin{eqnarray}
 \frac{m^2+\vec{k}_{\perp}^{2}}{x(1-x)}\leq \Lambda^2+4m^2
\,,\label{eq:domain}\end{eqnarray} 
which allows for states having a kinetic energy below the bare 
cut-off $\Lambda$.

\subsection{Dependence on the cut-off function}

Before discussing the dependence of the effective potential 
in Eq.(\ref{eq:56}) on the cut-off function $f$ one has to 
determine the dependence of $\Theta$ on its two energy 
arguments $D$, as given in Eq.(\ref{eq:ii77}). 
It is natural to assume that the similarity function $f(D;l)$ 
is a homogeneous function of its arguments
\begin{equation}
   f(D;l)=f(D^{\kappa}l)
\end{equation}
with some exponent $\kappa$. 
As examples we consider three types of similarity functions and the 
corresponding $\Theta$ functions:\\
the exponential cut-off 
\begin{equation}
   f(D;l)=\exp(-|D|l), \quad \kappa=1, \quad
   \Theta(D_e,D_{\bar e})=\frac{D_e}{D_e+D_{\bar e}}
,\end{equation}
the Gaussian cut-off
\begin{equation}
   f(D;l)=\exp(-D^2l), \quad \kappa=2, \quad
   \Theta(D_e,D_{\bar e})=\frac{D_e^2}{D_e^2+D_{\bar e}^2}
,\end{equation}
and the sharp cut-off
\begin{equation}
   f(D;l)=\theta(1/l-|D|^{\kappa})=\theta(1-|D|^{\kappa}l), \quad
   \Theta(D_e,D_{\bar e})=\theta(|D_e|-|D_{\bar e}|)
   =\theta(1-\frac{D_{\bar e}}{D_e})
.\end{equation}
In the last case $\kappa$ is an arbitrary positive number. 
In the first and in the last case we have used that $D_e$ 
and $D_{\bar{e}}$ have the same sign.
The second case corresponds to the solutions of the original 
flow equations, see Eqs.(\ref{eq:i19}) and (\ref{eq:f67}).
 
Assuming this homogeneity of $f$,  $\Theta$ is quite 
generally a function of the ratio of its two arguments since
\begin{equation}
   \Theta(D_e;D_{\bar e})=
   -\int_0^{\infty} dl' \frac{df(D_e^{\kappa}l')}{dl'}
   f(D_{\bar e}^{\kappa}l')
   =-\int_0^{\infty} dz \frac{df(z)}{dz}
   f\left(\left(\frac{D_{\bar e}}{D_e}\right)^{\kappa}z\right).
\end{equation}
Instead we may express it as a function of
\begin{eqnarray}
   \xi&=&\frac{D_e-D_{\bar e}}{D_e+D_{\bar e}}=\frac{\delta Q^2}{Q^2},
\\
   \Theta(D_e;D_{\bar{e}})&=&\frac 1 2 
   \left(1+\vartheta(\xi)\right)
.\end{eqnarray}
Due to Eq.(\ref{eq:ii18a}) $\vartheta(\xi)$ is an odd function, 
$\vartheta(-\xi)=-\vartheta(\xi)$.

Then the first expression in square brackets in (\ref{eq:56}) which contains
the singular part of the effective interaction reads
\begin{eqnarray}
   B_{\mu\nu}^{ex} &=& \left[
   g_{\mu\nu}\,
   \left(\frac{\Theta_{e \bar e}} {Q_{e}^2} +
         \frac{\Theta_{\bar e e}} {Q_{\bar e}^2}\right) + 
   \eta_\mu\eta_\nu\,\frac {\delta Q^2}{{q^+}^2} 
   \left(\frac{\Theta_{e \bar e}} {Q_{e}^2} - 
         \frac{\Theta_{\bar e e}} {Q_{\bar e}^2}\right)
	 \right]
\nonumber\\
   &=& \frac{g_{\mu\nu}}{Q^2} + 
   \frac{\xi^2-\xi\vartheta(\xi)}{1-\xi^2}
   \left(\frac{g_{\mu\nu}}{Q^2}-
   \frac{\eta_{\mu}\eta_{\nu}}{{q^+}^2}\right).
\label{eq:iii8}\end{eqnarray}
For the three similarity functions mentioned above one obtains
\begin{eqnarray}
   \mbox{exponential} & \vartheta(\xi)=\xi, 
   & \frac{\xi^2-\xi\vartheta(\xi)}{1-\xi^2}=0
, \label{eq:107}\\
   \mbox{Gaussian} & \vartheta(\xi) = \frac {2\xi} {1+\xi^2},
   & \frac{\xi^2-\xi\vartheta(\xi)}{1-\xi^2}=\frac{-\xi^2}{1+\xi^2}
, \label{eq:108}\\
   \mbox{sharp} & \vartheta(\xi)=\mbox{sign}(\xi),
   & \frac{\xi^2-\xi\vartheta(\xi)}{1-\xi^2}=\frac{-|\xi|}{1+|\xi|}
.\label{eq:109}\end{eqnarray}
We observe that the effective interaction depends explicitly
on the similarity function. 
The requirement of block diagonalization of the Hamiltonian 
determines the generator only up to a unitary transformation
of the blocks.
This explains why the effective interaction may depend 
on the similarity function.

We will discuss the dependence on this function further in
the next section, but mention that for the elimination
of the electron-phonon interaction in solid-state physics 
which yields the effective attractive interaction between 
electron pairs responsible for super-conductivity,
one may also choose different similarity functions,
see \cite{LeWe} and \cite{mie97,mie97a}.
For realistic spectra, Mielke \cite{mie97a} has found 
that the critical temperature calculated
from the Gaussian similarity function
and that suggested by Glazek and Wilson \cite{GlWi} 
differ by only 2\%,
the difference to those calculated by the conventional 
Eliashberg theory was only 5\%.

Since the kernel of the integral equation is manifestly 
frame-independent, one can evaluate it in the particular frame
$P^+=2m$ and $\vec{P}_{\perp}=0$. 
For the further discussions we choose to express the momenta as
\begin{eqnarray}
   \begin{array}{clcl} 
   p_1^+ &=m+p_\parallel, &\vec{p}_{1\perp} &=\phantom{-}\vec{p}_{\perp},\\
   p_2^+ &=m-p_\parallel, &\vec{p}_{2\perp} &=-\vec{p}_{\perp},
   \end{array}
\end{eqnarray}
and similarly for $p'_1$ and $p'_2$. 
With $q^+= q_\parallel = p'_\parallel - p_\parallel$,
$p^2\equiv\vec{p}_{\perp}^{\,2}+p_\parallel^{2}$ 
and  $q^2\equiv\vec{q}_{\perp}^{\,2}+q_\parallel^{2}$
the energy differences become 
\begin{eqnarray}
   D_e &=& 
   \frac {m^2+\vec {p'}_{\!\perp}^{\,2}} {m +p'_\parallel} -
   \frac {m^2+\vec {p }_{\!\perp}^{\,2}} {m +p _\parallel} -
   \frac {\vec {q}_{\!\perp}^{\,2}} {q^+} 
   = - 
   \frac {q^2} {q_\parallel} -
   \frac {{p }^2} {m-{p _\parallel}} +
   \frac {{p'}^2} {m-{p'_\parallel}}  
, \label{eq:ii110}\\
   D_{\bar e} &=& 
   \frac {m^2+\vec {p }_{\!\perp}^{\,2}} {m -p _\parallel} -
   \frac {m^2+\vec {p'}_{\!\perp}^{\,2}} {m -p'_\parallel} -
   \frac {\vec {q}_{\!\perp}^{\,2}} {q^+} 
   = -
   \frac {q^2} {q_\parallel} -
   \frac {{p'}^2} {m-{p'_\parallel}} +
   \frac {{p }^2} {m-{p _\parallel}} 
. \label{eq:ii111}\end{eqnarray}
The other quantities become correspondingly
\begin{eqnarray}
   D_e-D_{\bar e}&=& 2m \left(
   \frac {{p'}^2} {m^2-{p'_\parallel}^2} - 
   \frac {p^2} {m^2-{p_\parallel}^2}\right)
, \\
   \xi^2 &=& \left(\frac {\delta Q^2} {Q^2}\right)^2 = 
   \frac {q_\parallel^2} {Q^2}
   \frac {\left(D_e-D_{\bar e}\right)^2} {Q^2} 
, \\
   Q^2 &=& q^2 + q_\parallel \left(
   \frac {p_\parallel p^2} {m^2-{p_\parallel}^2} - 
   \frac {p'_\parallel{p'}^2} {m^2-{p'_\parallel}^2} \right) 
. \end{eqnarray}
Note that $-m \leq p_\parallel \leq m$ and that $p_\parallel$
{\em may not be } interpreted as the $z$-component of 
a single-particle momentum.

\section{Discussion and interpretation}
\label{sec:5}

The integral equation (\ref{eq:r96}) seems to have 
two kinds of singularities:
The `Coulomb singularities' $1/Q^2_{e}$ or $1/Q^2_{\bar e}$ 
and the `collinear singularity' $1/q_\parallel^2$.
Either denominator can become zero in the integral equation.
The Coulomb singularity is square-integrable and welcome since it 
provides the binding. The collinear singularity is disastrous.
If the coefficient of $1/q_\parallel^2$ is finite at 
$q_\parallel\rightarrow 0$ the integral equation is not solvable. 
The singularity structure of the
final result must therefore be discussed carefully.

First we observe, that in the case of the exponential 
similarity function the effective interaction 
between an electron and a positron becomes 
\begin{equation}
   U_{\rm eff} = - \frac{\alpha}{4\pi^2} 
   \frac {1} {\sqrt{x(1-x)x'(1-x')}} \left(
   \frac {<j^\mu J_\mu>_{ex}} {Q^2}
   + 
   \frac {<c^\mu C_\mu>_{an}} {M^2}
   \right)
\,.\label{eq:rr98}\end{equation}
The collinear singularity is wiped out since the coefficient
related to the similarity function vanishes, see Eq.(\ref{eq:107}).
This astoundingly compact formula exactly agrees with the 
Tamm-Dancoff approach \cite{TrPa}.
The explicit $x$-dependence in the denominator
of Eq.(\ref{eq:r94}) looks like the only remnant
of the light-cone formulation; all other quantities
are Lorentz-contracted scalars.

With the method of Hamiltonian flow one can calculate 
also the scattering amplitude for $e\bar e$-scattering  
\begin{equation}
   T_{\rm scattering} = - \frac{\alpha}{4\pi^2} 
   \frac {1} {\sqrt{x(1-x)x'(1-x')}} \left(
   \frac {<j^\mu J_\mu>_{ex}} {Q^2}
   + 
   \frac {<c^\mu C_\mu>_{an}} {M^2}
   \right)
\,,\end{equation}
see App.~\ref{app:a}. The expression agrees identically
with the Feynman amplitude.
In the scattering process the free 4-momentum $P_\mu$
is conserved, and thus the momentum transfer along the 
electron and the positron line are the same, {\it i.e.}
$Q_{e}^2 = Q_{\bar e}^2 = Q^2$. Therefore $\delta Q^2=0$
and thus $\xi=0$. According to Eq.(\ref{eq:iii8}),
the coefficient of the collinear singularity vanishes
identically. 

The scattering amplitude can be investigated also in 
light-cone perturbation theory \cite{LeBr}.
Its exchange part is presented in all details in 
Sec.~3.4 of \cite{bpp97}. As can be pursued there, 
the collinear singularity appears in both 
the instantaneous interaction and the second order
amplitude $VGV$ of the vertex interaction,
however such that the two contributions 
{\em cancel each other exactly}. 
Only the integrable Coulomb singularity remains.
One concludes that Hamiltonian flow equations
are governed {\em de facto} by the same mechanism 
of cancelation but that the disappearance of the
collinear singularity is somewhat more subtle.
To understand that better 
two approximations are discussed in the sequel.

First, let us expand the kernel up to terms linear in $p/m$. 
To this order holds 
\begin{equation}
   {\bar u}(p',\lambda')\gamma^{\mu} u(p,\lambda)=
   \left\{ \begin{array}{c c} 
   2m\delta^{\lambda'}_{\lambda}, & \mu=+,-, \\
   0, & \mu=1,2. \end{array} \right.
\end{equation}
The bracket symbols of Eq.(\ref{eq:r94}) become then
\begin{equation}
   < \gamma^{\mu} \gamma^{\nu} > g_{\mu\nu} = 
   < \gamma^{\mu} \gamma^{\nu} > \eta_{\mu} \eta_{\nu} = 16m^2 
   \delta^{\lambda'_1}_{\lambda_1} 
   \delta^{\lambda'_2}_{\lambda_2}.
.\label{eq:i118}\end{equation}
According to Eqs.(\ref{eq:ii110}) and (\ref{eq:ii111})
one has to this order 
\begin{equation}
   D_{e} = D_{\bar e} = - {q^2}/{q_\parallel} 
.\end{equation}
Consequently, $\delta Q^2=0$ and that alone is sufficient to
wipe out all dependence on the similarity function.
The exchange part becomes
\begin{equation}
   U_{\rm eff} = - \frac{\alpha}{\pi^2} 
   \frac{1}{q^{2}}\ (2m)^2
   \delta^{\lambda'_1}_{\lambda_1} \delta^{\lambda'_2}_{\lambda_2} 
\,.\end{equation}
The factor $(2m)^2$ is a light-cone peculiarity,
see Sec.~4.9 of \cite{bpp97}, and the factor in front of it 
is precisely the Fourier transform of the 
familiar Coulomb potential in three space-dimensions.
Hence the effective electron-positron
interaction Eq.(\ref{eq:56}) is bound to produce
the Bohr spectrum.

Next, expand the kernel up to second order in $p^2/m^2$.
This  gives the familiar Breit-Fermi spin-spin and 
tensor interactions \cite{BrPe}, that insures the correct 
spin-splittings for the positronium ground state 
and restores the rotational invariance \cite{JoPeGl}.
Brisudova {\it et al.} \cite{BrPe} state also that the 
`$\eta_{\mu}\eta_{\nu}$' term may influence the spin-orbit 
coupling in second-order bound-state perturbation theory.
The correct singlet-triplet splitting is observed also
in the numerical solutions of the integral equation 
\cite{TrPa,kpw92} with the effective interaction 
(\ref{eq:rr98}) as well as with flow equations (see further).

Let us now consider the case of a general similarity function.
In order to do this we have to discuss the expressions of the previous 
subsection. We realize, that as the momentum transfer 
$q=p'-p$ tends to zero $D_e-D_{\bar e}$ tends linearly to zero and $Q^2$
vanishes quadratically in $q$. This approach to zero is anisotropic.
Only at $p=0$ itself $Q^2$ approaches 0 isotropically and the linear
contribution in $D_e-D_{\bar e}$ disappears. One can also show, that $Q^2$
can only vanish, if $q=0$. It is not sufficient, that $q^+$ vanishes.
Consequently $\xi$ is always finite (which also follows from the observation,
that $Q_e^2$ and $Q_{\bar e}^2$ are positive and thus $-1\le \xi \le +1$). 
However as $q$ vanishes
$\xi$ will in general depend on the direction from which $q$ approaches zero.
Moreover we realize, that for sufficiently smooth similarity functions
$f$ the function $\vartheta(\xi)$ will be analytic as in the cases of the 
exponential and Gaussian cut-off. Then the pre-factor 
$(\xi^2-\xi\vartheta(\xi))/(1-\xi^2)$ in (\ref{eq:iii8}) contains a factor
$\xi^2$, which itself contains a factor ${q^+}^2$, which cancels against
the denominator ${q^+}^2$ of the $\eta_{\mu}\eta_{\nu}$ term. Thus the 
interaction becomes only singular if $q$ approaches 0 where it diverges like
$1/q^2$. This is however not true for the sharp cut-off, where only
one factor $q^+$ can be cancelled and one $q^+$ remains in the denominator
of the $\eta_{\mu}\eta_{\nu}$ term. Thus for a smooth cut-off one gets rid
of the collinear singularity.

We realize further that $\xi$ is of order $p/m$. 
Since however the Bohr momentum is of order $m\alpha$ the contribution 
due to the second term in (\ref{eq:iii8})
is smaller by a factor $\alpha^2$ in comparison to the leading term
$g_{\mu\nu}/Q^2$.

Thus we realize that in this order we have in addition to the leading term
a contribution
\begin{equation}
   -\frac{\alpha}{4\pi^2} 
   < \gamma^{\mu} \gamma^{\nu} >_{ex} (1-\vartheta'(0))
   \xi^2 \left(\frac{g_{\mu\nu}}{Q^2}
   -\frac{\eta_{\mu}\eta_{\nu}}{{q^+}^2}\right)
\label{eqiii:22a}\end{equation}
to the interaction. It obviously depends on the similarity function
via the derivative $\vartheta'(0)$. 
We will shortly discuss the leading contribution 
and use Eq.(\ref{eq:i118}).
Collecting terms one obtains 
\begin{equation}
   (1-\vartheta'(0))\frac{16\alpha}{\pi^2} 
   \frac{q_\perp^2(p'^2-p^2)^2}{q^6}
\label{eqiii:23}\end{equation}
as the contribution to the effective interaction. 
This interaction is spin-independent but 
anisotropic. It depends on the similarity function and is of order $\alpha^2$
in comparison to that of the leading Coulomb interaction. We emphasize
as already pointed out in \cite{GuWe}, that in this order in $\alpha^2$
also one- and two-loop terms contribute. Since they will also in general
depend on the cut-off chosen, there should be a cancelation between the
term found here and loop terms. It was argued there, that in order $\alpha^2$
the one- and two-loop terms only contribute spin-independent interactions
and interactions of one spin (spin-orbit coupling), but not interactions 
between
both spins (spin-spin and tensor interaction), which is in agreement with our
finding that the interaction (\ref{eqiii:23}) does not depend on the spin.

\section{Numerical solution for positronium spectrum}
\label{sec:6}

We solve the integral equation Eq.(\ref{eq:r96}),
with interaction kernel given in Eq.(\ref{eq:56}), 
for positronium mass spectrum numerically.
Effective interaction with different choice of cut-offs
is sumarized in Appendix (\ref{app:b}).

\subsection{Formulation of the problem}
\label{subsec:6.1}

In polar coordinates the light-front variables are
$(\vec k_{\perp};x)=(k_{\perp},\varphi;x)$;
therefore the matrix elements of the effective interaction Eq.(\ref{eq:56})
depend on the angles $\varphi$ and $\varphi^{\prime}$, i.e. 
$\langle x,k_{\perp},\varphi;\lambda_1,\lambda_2|V_{\rm eff}|
x',k'_{\perp},\varphi';\lambda'_1,\lambda'_2\rangle$.
In order to introduce the spectroscopic notation for positronium
mass spectrum we integrate out the angular degree of freedom,
$\varphi$, introducing a discrete quantum number
$J_z=n$, $n\in {\bf Z}$
(actually for the annihilation channel only $|J_z|\leq 1$ is possible),
\begin{eqnarray}
&&  \hspace{-1.5cm} 
      \langle x, k_{\perp}; J_z, \lambda_1, \lambda_2
      |\tilde{V}_{\rm eff}|
       x',k'_{\perp};J'_z,\lambda'_1,\lambda'_2\rangle
\nonumber\\
&=& \frac{1}{2\pi}\int_0^{2\pi}d\varphi {\rm e}^{-iL_z\varphi}
                  \int_0^{2\pi}d\varphi'{\rm e}^{iL'_z\varphi'}
      \langle x, k_{\perp}, \varphi; \lambda_1, \lambda_2
      |V_{\rm eff}|
       x',k'_{\perp},\varphi';\lambda'_1,\lambda'_2\rangle
\nonumber\\
&&
\,\label{eq:r43}\end{eqnarray}
where $L_z=J_z-S_z$; $S_z=\frac{\lambda_1}{2}+\frac{\lambda_2}{2}$ 
and the states
can be classified (strictly speaking only for rotationally invariant
systems) according to their quantum numbers of total angular momentum $J$,
orbit angular momentum $L$, and total spin $S$. 
Definition of angular momentum operators in light-front
dynamics is problematic because they include interactions.

The matrix elements of the effective interaction
before integrating over the angles,
$ \langle x, k_{\perp}, \varphi; \lambda_1, \lambda_2
|V_{\rm eff}| x',k'_{\perp},\varphi';\lambda'_1,\lambda'_2\rangle$,
and after the integration inroducing the total momentum, $J_z$,
$ \langle x, k_{\perp}; J_z, \lambda_1, \lambda_2|\tilde{V}_{\rm eff}|
   x',k'_{\perp};J'_z,\lambda'_1,\lambda'_2\rangle$
for different cut-off functions are given in the exchange and 
annihilation channels in Appendices (\ref{app:c}) and (\ref{app:d}), respectively.

Now we proceed to solve for the positronium spectrum 
in all sectors of $J_z$. For this purpose we formulate 
the light-front integral equation Eq.~(\ref{eq:r96}) in the form
where the integral kernel is given by the effective interaction
for the total momentum $J_z$, Eq.~(\ref{eq:r43}). 
We introduce instead of Jacobi momentum 
$(x,\vec{k}_{\perp})$ the three momentum in the center of mass frame 
$\vec{p}=(p_z,\vec{k}_{\perp})=
(\mu\cos\theta,\mu\sin\theta\cos\varphi,\mu\sin\theta\sin\varphi)$
as follows
\begin{eqnarray}
   x = \frac{1}{2}\left(1+\frac{p_z}{\sqrt{{\vec p}^{\, 2}+m^2}} \right)
\,,\label{eq:eq3}\end{eqnarray}
where the Jacobian of this transformation $dx/dp_z$ is
\begin{eqnarray}
&& J=\frac{1}{2}
\frac{m^2+\mu^2 \sin^2\theta}{(m^2+\mu^2)^{3/2}}
\,.\label{r47}\end{eqnarray}

One obtaines then the integral equation
\begin{eqnarray}
  && (M_n^2-4(m^2+\mu^2))
   \tilde{\psi}_n(\mu,\cos\theta;J_z, \lambda_1, \lambda_2)
\nonumber\\
  &+&   \sum_{J'_z,\lambda'_1,\lambda'_2}
        \int_{D}d\mu'\int_{-1}^{+1}d\cos\theta'
        \frac{\mu^{'2}}{2}
        \frac{m^2+\mu^{'2}(1-\cos^2\theta')}{(m^2+\mu^{'2})^{3/2}}
\nonumber\\
 &\times& \langle\mu, \cos\theta; J_z, \lambda_1, \lambda_2
         |\tilde{V}_{\rm eff}|
               \mu',\cos\theta';J'_z,\lambda'_1,\lambda'_2\rangle
 \tilde{\psi}_n
        (\mu',\cos\theta';J'_z,\lambda'_1,\lambda'_2)=0
\,.\label{eq:r48}\end{eqnarray}
The integration domain $D$, defined in Eq.~(\ref{eq:domain}), 
is given now by $\mu\in [0;\frac{\Lambda}{2}]$.
Neither $L_z$ nor $S_z$ are good quantum numbers; therefore
we set $L_z=J_z-S_z$.

The integral equation Eq.~(\ref{eq:r48}) is used to calculate
positronium mass spectrum numerically.
Note, that if one succeeds to integrate out the angular degrees of freedom
for the effective interaction Eq.~(\ref{eq:r43}) analytically,
one has $2$-dimensional integration in Eq.~(\ref{eq:r48})
instead of $3$-dimensional one in the original
integral equation~(\ref{eq:r96}) to perform numerically. 

We use the numerical code \cite{TrPa2}, worked out by Uwe Trittmann  
for the similar problem \cite{TrPa}. 
This code includes for the numerical integration
the Gauss-Legendre algorithm (Gaussian quadratures). 
To improve the numerical convergence
the technique of Coulomb counterterms is included.
The problem has been solved for all components of the total
angular momentum, $J_z$. 

Positronium spectrum is mainly defined by the Coulomb singularity
\begin{eqnarray}
 \vec q \rightarrow 0
\,,\label{eq:limit1}\end{eqnarray}
which is an integrable one analytically and also, by use of technique 
of Coulomb counterterms, numerically. 
In this region the effective interaction Eq.(\ref{eq:56})
has leading Coulomb behavior  Eq.(\ref{eq:rr98}),
independent on the cut-off function.
We use therefore in numerical procedure standard Coulomb counterterms,
introduced for the Coulomb problem Eq.(\ref{eq:rr98}) \cite{TrPa,TrPa2},
for all cut-offs.
 
Also we expect therefore
the same pattern of levels for different cut-offs,
that is proved numerically further.

Another important limiting case to study effective interaction 
Eq.(\ref{eq:56}), namely its exchange part Eq.(\ref{eq:iii8}), 
is the collinear limit
\begin{eqnarray}
  q^+ \rightarrow 0
\,,\label{eq:limit2}\end{eqnarray}  
that is special for light-front calculations.
Exchange part of the effective interaction is given by  Eq.(\ref{eqiii:22a}),
which is finite in this limit. This is true for 
the regular cut-off functions, as in the case of exponential 
and gaussian cut-offs, where the derivative 
$d\vartheta(0)/d\xi$ is well defined. For sharp cut-off
this condition is not fulfilled, and the effective interaction
contains the $1/q^+$ type of singularity 
(see Appendix (\ref{app:b}). We do not associate any physics with this singularity,
and consider it as a consequence of artificial choice of cut-off,
which corresponds to singular generator of unitary transformation
Eq.(\ref{eq:55a}). In numerical calculations we omit $'\eta_\mu\eta_\nu'$ term
in exchange channel for sharp cut-off.

\begin{table}
\begin{tabular}{c||c||c||c||c||c||c}
$n$ & Term & $B_{ETPT}$ & $B_E$ & $B_G^{\eta}$& $B_G$ & $B_S$ \cr \hline \hline
1   & $1^1S_0$ & 1.118125   & 1.049550 & 1.101027 & 1.026170 & 0.920921 \\
2   & $1^3S_1$ & 0.998125   & 1.001010 & 1.049700 & 0.981969 & 0.885347 \\
3   & $2^1S_0$ & 0.268633   & 0.260237 & 0.266490 & 0.260642 & 0.242607 \\
4   & $2^3S_1$ & 0.253633   & 0.253804 & 0.259506 & 0.254765 & 0.234312 \\
5   & $2^1P_1$ & 0.253633   & 0.257969 & 0.263056 & 0.257664 & 0.237611 \\
6   & $2^3P_0$ & 0.261133   & 0.267070 & 0.273826 & 0.266563 & 0.243075 \\
7   & $2^3P_1$ & 0.255508   & 0.259667 & 0.265412 & 0.260127 & 0.238135 \\
8   & $2^3P_2$ & 0.251008   & 0.255258 & 0.260345 & 0.255498 & 0.236383 
\end{tabular}
\caption{Binding coefficients, $B_n=4 (2-M_n)/\alpha^2$
($\alpha=0.3$), for the lowest modes of the positronium spectrum 
at $J_z=0$ for the equal time perturbation theory up to order 
$\alpha^4$ ($B_{ETPT}$ \cite{TrPa}) compared to our
calculations with exponential ($B_E$), Gaussian ($B_G$) and sharp
($B_S$) cut-offs. $B_G$ is obtained using only $g_{\mu\nu}$ part
of interaction; for $B_G^{\eta}$ $'\eta_\mu\eta_\nu`$ term is included.   
Exchange channel is considered.}
\label{tab:1}
\end{table}

\begin{table}
\begin{tabular}{c||c||c||c||c||c}
$n$ & Term     & $B_E$    & $B_G^{\eta}$& $B_G$    & $B_S$    \cr \hline \hline
1   & $1^1S_0$ & 1.049550 & 1.101270 & 1.026170 & 0.920921 \\
2   & $1^3S_1$ & 0.936800 & 0.978018 & 0.921847 & 0.834004 \\
3   & $2^1S_0$ & 0.260237 & 0.266490 & 0.260642 & 0.242624 \\
4   & $2^3S_1$ & 0.255292 & 0.260383 & 0.255615 & 0.234338 \\
5   & $2^1P_1$ & 0.257969 & 0.263056 & 0.257664 & 0.236383 \\
6   & $2^3P_0$ & 0.267090 & 0.273847 & 0.266626 & 0.243075 \\
7   & $2^3P_1$ & 0.259667 & 0.265412 & 0.260127 & 0.237611 \\
8   & $2^3P_2$ & 0.245615 & 0.250821 & 0.247091 & 0.230901  
\end{tabular}
\caption{Binding coefficients, $B_n=4 (2-M_n)/\alpha^2$
($\alpha=0.3$), for the lowest modes of the positronium spectrum 
at $J_z=0$ for our calculations with exponential ($B_E$), 
Gaussian ($B_G$) and sharp ($B_S$) cut-offs. 
$B_G^{\eta}$ includes $'\eta_\mu\eta_\nu`$ term
in exchange channel; $B_G$ does not. 
Exchange and annihilation channels are considered.}
\label{tab:2}
\end{table}

\begin{table}
\begin{tabular}{c|| c|| c|| c|| c}
$n$ &Term & $\delta B_E$ & $\delta B_G$ & $\delta B_S$ \\ \hline \hline\\[-8pt]
2   & $1^3S_1$ &  6.30 $10^{-4}$  &  1.76 $10^{-3}$  & 1.18 $10^{-3}$ \\
4   & $2^3S_1$ &  8.40 $10^{-5}$  &  1.77 $10^{-4}$  & 9.0 $10^{-5}$ \\
5   & $2^1P_1$ & -1.30 $10^{-5}$  & -7.47 $10^{-4}$  &-9.1 $10^{-5}$ \\
7   & $2^3P_1$ & -4.08 $10^{-4}$  & -4.08 $10^{-4}$  & 1.4 $10^{-4}$ \\
8   & $2^3P_2$ &  5    $10^{-6}$  & -7.7 $10^{-5}$   & 4.15 $10^{-4}$ 
\end{tabular}
\caption{Difference in the corresponding energy levels between $J_z\!=\!0$
and $J_z\!=\!1$ states for exponential ($\delta B_E$), 
Gaussian ($\delta B_G$) and sharp ($\delta B_S$) cut-offs.
Exchange channel is considered.}
\label{tab:3}
\end{table}

\begin{table}
\begin{tabular}{c|| c|| c|| c|| c}
$n$ & Term & $\delta B_E$ & $\delta B_G$ & $\delta B_S$ \\ \hline \hline\\[-8pt]
2   & $1^3S_1$ & -1.411 $10^{-3}$ & -7.86 $10^{-4}$  &-1.65 $10^{-3}$ \\
4   & $2^3S_1$ & -4.1 $10^{-5}$   & -4.0 $10^{-5}$   &-1.15 $10^{-4}$ \\
5   & $2^1P_1$ & -6.4 $10^{-5}$   & -6.52 $10^{-4}$  &-4.60 $10^{-4}$ \\
7   & $2^3P_1$ & -4.69 $10^{-4}$  & -4.74 $10^{-4}$  &-1.40 $10^{-4}$ \\
8   & $2^3P_2$ & -1.96 $10^{-4}$  & -1.36 $10^{-4}$  &-2.44 $10^{-4}$ 
\end{tabular}
\caption{Difference in the corresponding energy levels between $J_z\!=\!0$
and $J_z\!=\!1$ states for exponential ($\delta B_E$), 
Gaussian ($\delta B_G$) and sharp ($\delta B_S$) cut-offs.
Exchange and annihilation channels are considered.}
\label{tab:4}
\end{table}

\subsection{Discussion of numerical results}
\label{subsec:6.2}

\begin{figure}[thbp]
\centerline{\epsfxsize=\textwidth \epsfbox{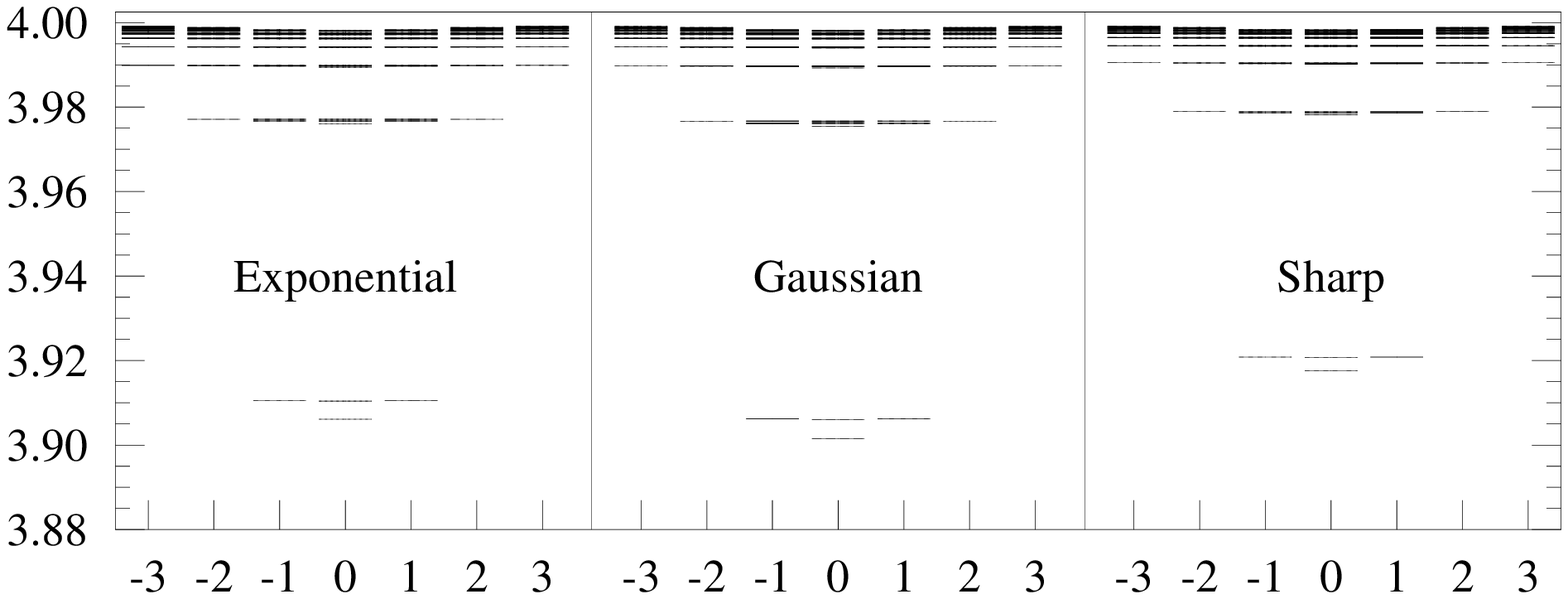}}
\caption{The invariant mass-squared spectrum $M_i^2$ for positronium 
   versus the projection of the total spin, $J_z$, excluding
   annihilation with exponential, Gaussian and sharp cut-offs.
   The number of integration points is $N_1=N_2=21$.}
\label{fig:5}
\end{figure}

\begin{figure}[thbp]
\centerline{\epsfxsize=\textwidth \epsfbox{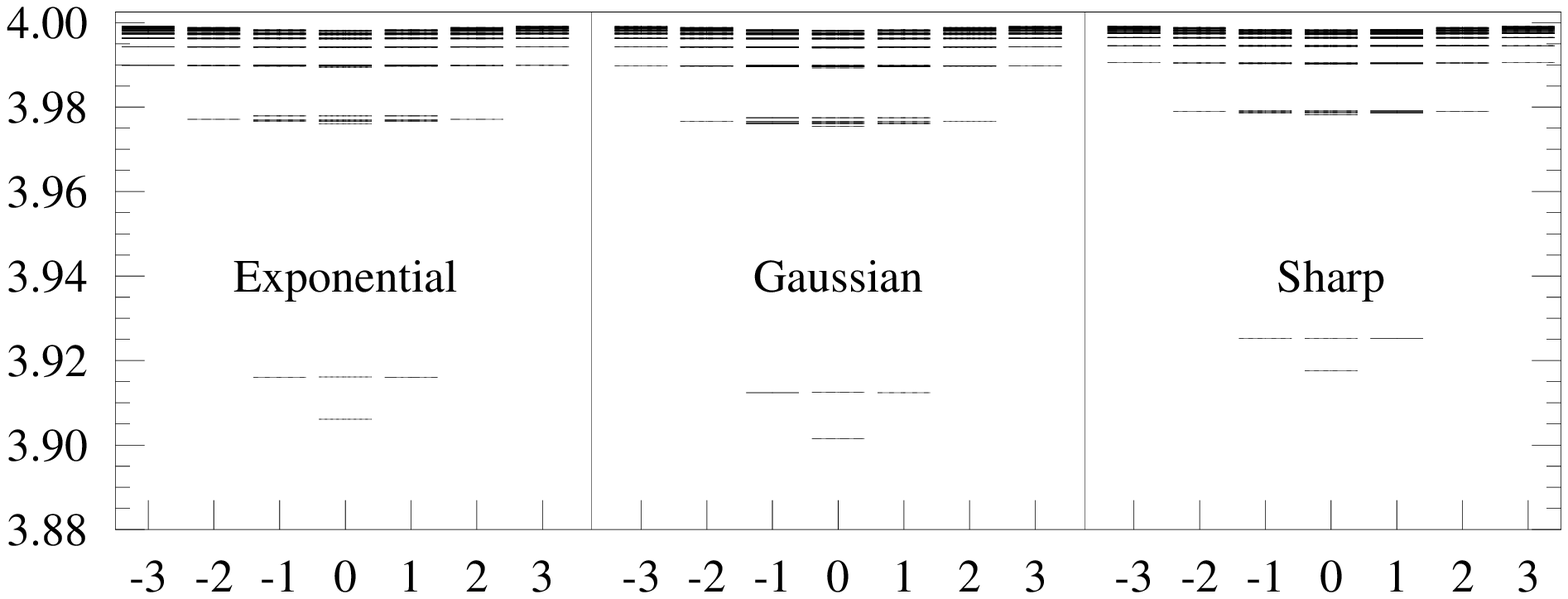}}
\caption{The invariant mass-squared spectrum $M_i^2$ for positronium 
   versus the projection of the total spin, $J_z$, including
   annihilation with exponential, Gaussian and sharp cut-offs.
   The number of integration points is $N_1=N_2=21$.}
\label{fig:6}
\end{figure}

We place the results of calculations for three different cut-offs,
performed in exchange and including both exchange and annihilation
channels, in Tables (\ref{tab:1}) and (\ref{tab:2}), respectively.   
The corresponding set of figures is presented in Fig.(\ref{fig:5})
and Fig.(\ref{fig:6}).
We get the ionization threshold at $M^2\sim 4m^2$,
the Bohr spectrum, and the fine structure.
Including annihilation part increases the splittings twice as large
for the lowest multiplets.

We argued that the region of Coulomb singularity,
and hence $'g_{\mu\nu}'$ part of effective interaction,
determines mainly the positronium spectrum. However,
including $'\eta_\mu\eta_\nu'$ part for gaussian cut-off
shifts spectrum as a whole down to about $5-7\%$,
since this part is diagonal in spin space (Appendix (\ref{app:c})),
and improves the data to be near the result obtained in covariant
equal time calculations (Table(\ref{tab:1})). 
For the sharp cut-off the lowest multiplet is placed higher than the one 
in case of exponential and gaussian cut-offs. The reason is in disregarding
the infrared divergent $'\eta_\mu\eta_\nu'$ part.
Presumably,
it is necessary to take into account $'\eta_\mu\eta_\nu'$ term
in exchange channel also for sharp cut-off
after the proper regularization of infrared longitudinal
divergences is done.

As one can see from presents figures, certain mass eigenvalues at 
$J_z=0$ are degenerate with certain eigenvalues at other $J_z$
to a very high degree of numerical precision. As an example,
consider the second lowest eigenvalue for $J_z=0$.
It is degenerate with the lowest eigenvalue for $J_z=\pm 1$,
and can thus be classified as a member of the triplet with $J=1$.
Correspondingly, the lowest eigenvalue for $J_z=0$ having no companion
can be classified as the singlet state with $J=0$.
Quite in general one can interpret degenerate multiplets 
as members of a state with total angular momentum
$J=2J_{z,max}+1$. One can get the quantum number
of total angular momentum $J$ from the number of degenerate states
for a fixed eigenvalue $M_n^2$. One can make
contact with the conventional classification scheme 
$^{2S+1}L_{J}^{J_z}$, as indicated in Tables (\ref{tab:1})-(\ref{tab:2}).

Such pattern of spectrum is driven by rotational invariance.
To trace rotational symmetry we calculate the difference
of energy levels between $J_z=0$ and $J_z=1$ states
for the lowest multiplets. The data are given for exchange and
including annihilation channnel in Tables (\ref{tab:3}) and (\ref{tab:4}), 
respectively.
Including annihilation channel improves the extent of degeneracy
(see Table(\ref{tab:4}) and Figure(\ref{fig:6})).

Concerning the spin-splittings the best agreement with covariant calculations 
is obtained for gaussian cut-off, the worst results are for sharp cut-off.
Rotational invariance is traced on the level of spectrum by studing
the degree of degeneracy of corresponding states with the same total momentum
but different projection $J_z$ in the multiplet. Again, better results
are obtained for exponential and gaussian cut-off functions than for sharp cut-off.
This suggests, that smooth cut-off functions are preferable to perform calculations. 

Generally, the impact of the different choice of cut-off functions
on the spectrum is small.

In this work we solve the bound state integral equation for the one fixed
integration interval. Integration domain introduces the ultraviolet cut-off
dependence of invariant mass squared $M^2(\Lambda)$, that reflects
renormalization group properties of the effective coupling constant.
We leave this question for the future study.

\newpage

\section{Summary and conclusions}
\label{sec:7} 

We have applied the method of Hamiltonian flow to the
canonical Hamiltonian of quantum electrodynamics in the front 
form and derived an effective interaction for an
electron and a positron, which acts only in the $e\bar e$-sector 
of Fock space. To lowest order of approximation is the
familiar Coulomb interaction.
In this first of a series of papers we have 
restricted ourselves to include terms up to the second order in the
coupling constant $e$. By reasons of simplicity (almost) 
all aspects of renormalization theory within a Hamiltonian approach
have been disregarded in this first assault. 
Depending on the particular choice of the similarity
function one gets perfect agreement with other approaches particularly
with the method of iterated resolvents. 
Special emphasis is put on the impact of the collinear singularity.
Depending on the similarity function, the collinear singularity
is either absent from the outset,  
or it is shown explicitly that it has no impact on the solubility 
of the final integral equation.

The numerical solution of positronium bound state problem,
with the effective electron positron interaction obtained 
by the flow equations,
is presented. No approximations along numerical procedure are done.
 
One concludes that the method of Hamiltonian flow equations
looks like an excellent tool to progress further, particularly
to attack the severe problems of renormalization theory
within a non-perturbative Hamiltonian approach to field theory
and thus to apply it eventually to non-abelian gauge field theory
and quantum chromodynamics.

\noindent 
{\bf Acknowledgements}

One of us (E.G.) is grateful to Stan Glazek and Tim Walhout
for many useful discussions. Also E.G. thanks Martin Schaden 
for helpful suggestions. The work of E.G. was supported
by 'Fortbildungsstipendium (in Anlehnung an die 
F\"orderrichtlinien) der Max-Planck-Gesellschaft'.
The work of G.P. was partly supported by the US DOE grant DE-FG02-86ER40251
and Hungarian grant OTKA F026622.

\newpage
\appendix
\section{Flow equations and Feynman amplitudes}
\label{app:a} 

There must be a connection between the Feynman diagrammatic
technique which is based on the action
and the flow equations which are based on the Hamiltonian.
It will be studied here up to second order in the coupling
constant

The difference resides in the $l$-ordering of the generator 
$\eta(l)$ in the unitary operator  
\begin{equation}
   U(l)=T_l {\rm exp}\, \left(\int_0^l \eta(l')dl'\right)
\,.\end{equation}
The transformed Hamiltonian reads
\begin{equation}
   H(l)=U(l)H(0)U^+(l)
\,.\end{equation}
With the full unitary transformation
\begin{equation}
   U(\infty)={\rm e}^{-S}
\end{equation}
one has thus up to
the second order in the coupling constant
\begin{equation}
   H(\infty)={\rm e}^{-S}H(0){\rm e}^{S}
   =H(0)+[H(0),S]+[[H(0),S],S]+...
\,,\end{equation}
and
\begin{equation}
   S=S^{(1)}+S^{(2)}+... = - 
   \int_0^{\infty}\!\!dl\ \eta(l) - 
   \frac{1}{2}\int_0^{\infty}\!\!dl
   \int_0^{l}\!\!dl'\ [\eta(l),\eta(l')]+...
\,.\end{equation}
The series for the effective Hamiltonian reads
\begin{eqnarray}
   H(\infty) &=& H_0(0)+V(0)+[H_0(0),S^{(1)}]+[V(0),S^{(1)}]
\nonumber\\
   &+&\frac{1}{2}[[H_0(0),S^{(1)}],S^{(1)}]+[H_0(0),S^{(2)}]+...
\,.\end{eqnarray} 
The first order term vanishes
\begin{equation}
   V(0)+[H_0(0),S^{(1)}]=0
\,,\end{equation}
which gives rise to
\begin{equation}
   H(\infty) = H_0(0)+\frac{1}{2}[V(0),S^{(1)}]+[H_0(0),S^{(2)}]
\,.\label{eq:112}\end{equation}
Denoting by $S^{(i)}$ the order with respect to $e$
and omitting the instantaneous terms gives
\begin{eqnarray}
   S^{(1)} &=& {e\over\sqrt{(2\pi)^3}}
   \int\!{[d^3p_1]\over\sqrt{2p^+_1}} 
   \int\!{[d^3p_2]\over\sqrt{2p^+_2}} 
   \int\!{[d^3p_3]\over\sqrt{2p^+_3}}
   \,\delta^{(3)}(p_{1} - p_{2} - p_{3}) \,
   \frac {1} {D(p_1,p_2,p_3)}
\nonumber\\
   &\times&  
   \left[ b^\dagger_1 b_2 a_3 \, 
   (\overline u_1 \slash\!\!\!\epsilon_3 u_2) 
   -d_1^\dagger d _2 a_3 \, 
   (\overline v_2 \slash\!\!\!\epsilon_3 v_1)
   +a_1^\dagger d_2 b _3 \, 
   (\overline v_2 \slash\!\!\!\epsilon_1^\star u_3) 
   \right] - h.c.
\,.\end{eqnarray} 
and
\begin{eqnarray}
    S^{(2)} &=& {e^2\over(2\pi)^3}
    \sum_{\lambda_1,\lambda_2,\lambda'_1,\lambda'_2}
   \int\!{[d^3p_1]\over\sqrt{2p^+_1}}
   \int\!{[d^3p_2]\over\sqrt{2p^+_2}} 
   \int\!{[d^3p_1^\prime]\over\sqrt{2p^{\prime+}_1}} 
   \int\!{[d^3p_2^\prime]\over\sqrt{2p^{\prime+}_2}} 
\nonumber\\
   &\times&
   \delta^{(3)}(p_1 + p_2 - p'_1 - p'_2) 
   \,\frac {d_{\mu\nu}(q)} {q^+}   
   \left( -\frac{1}{2D_1D_2}\right)
   \left(\Theta(D_1,D_2)-\Theta(D_2,D_1)\right)
\nonumber\\
   &\times&
   {(\overline u(p_1,\lambda_1)\gamma^\mu u(p'_1,\lambda'_1))
   \over\sqrt{2p_1^{+}}\ \sqrt{2p_1^{\prime+}} }
   {(\overline v(p'_2,\lambda'_2)\gamma^{\nu} v(p_2,\lambda_2)) 
   \over\sqrt{2p_2^{\prime+}}\ \sqrt{2p_2^{+}}}
 \nonumber\\
   &\times& 
   \ :b^\dagger (p_1,\lambda_1) b(p'_1,\lambda'_1)
   \, d^\dagger (p_2,\lambda_2) d(p'_2,\lambda'_2) : 
\,.\end{eqnarray} 
The $D$ are of course the energy differences defined in
Eq.(\ref{eq:40a}), particularly
$D_1={p'_1}^- - p_1^{-} - (p'_1-p_1)^-$ and 
$D_2={p'_2}^- - p_2^{-} - (p_2-p'_2)^-$, 
and $q$ denotes the 4-momentum of the exchanged gluon. 
The matrix elements in the $e\bar e$-sector become correspondingly
\begin{eqnarray}
   <e\bar{e}|[V(0),S^{(1)}]|e'\bar{e'}> &=& -\frac{e^2}{2(2\pi)^3}
   \frac{(\overline u(p_1,\lambda_1)\gamma^\mu
   u(p'_1,\lambda'_1))}{\sqrt{2p_1^+}\sqrt{2p_1^{' +}}}
   \frac{(\overline v(p'_2,\lambda'_2)\gamma^\nu
   v(p_2,\lambda_2))}{\sqrt{2p_2^{'+}}\sqrt{2p_2^+}}
\nonumber\\
   &\times& \frac{d_{\mu\nu}(q)}{q^+}
   \left(\frac{1}{D_1}+\frac{1}{D_2}\right)
\label{eq:117}\end{eqnarray}
and
\begin{eqnarray}
   <e\bar{e}|[H_0(0),S^{(2)}]|e'\bar{e'}> &=& -\frac{e^2}{2(2\pi)^3}
   \frac{(\overline u(p_1,\lambda_1)\gamma^\mu
   u(p'_1,\lambda'_1))}{\sqrt{2p_1^+}\sqrt{2p_1^{' +}}}
   \frac{(\overline v(p'_2,\lambda'_2)\gamma^\nu
   v(p_2,\lambda_2))}{\sqrt{2p_2^{'+}}\sqrt{2p_2^+}}
\nonumber\\
   &\times& 
   \frac{d_{\mu\nu}(q)}{q^+}
   \left(-\frac{1}{2}\frac{D_1-D_2}{D_1D_2}\right)
   \left(\Theta(D_1,D_2)-\Theta(D_2,D_1)\right)
\,.\label{eq:118}\end{eqnarray} 
Inserting the latter two equation into Eq.(\ref{eq:112})
gives the generated interaction in Eq.(\ref{eq:ii81}).
Note that the $\Theta$-factor in Eq.(\ref{eq:ii81})
satisfies
\begin{equation}
   \frac{\Theta_{12}} {D_1}+
   \frac{\Theta_{21}}{D_2} = 
   \frac {1} {2} 
   \left(\frac {1} {D_1} + \frac {1} {D_2} \right) +
   \frac {1} {2} 
   \left(\frac {1} {D_1} - \frac {1} {D_2} \right) 
   \left(\Theta_{12}-\Theta_{21}\right)
\,.\label{eq:ii18b}\end{equation}
The first term in Eq.(\ref{eq:ii81}) corresponds 
therefore to the result of perturbation
theory and comes from Eq.(\ref{eq:117}). The second term 
originates from the $l$-ordering (\ref{eq:118}) 
and vanishes on the mass shell.

\newpage
\section{Defining different cut-offs}
\label{app:b}

In this appendix we summarize the results for 
the effective electron-positron interaction, generated 
by the flow equations with different similarity functions. 
In the practical work, three different similarity function
will be studied explicitly:
 
(1) the exponential cut-off, 
(2) the gaussian cut-off, and
(3) the sharp cut-off.

\noindent
(1) {\bf Exponential cut-off} 
\begin{eqnarray}
 f(D;l) &=& {\rm exp}\left(- \vert D \vert l \right)
\nonumber\\
 \Theta(D_e,D_{\bar{e}}) &=& \frac{D_e}{D_e+D_{\bar{e}}}
 ~;~ \vartheta(\xi) = \xi
\nonumber\\
 V_{\rm eff} &=& \frac{\alpha}{4\pi^2}
        \langle\gamma^{\mu}\gamma^{\nu}\rangle_{ex}
         \frac{g_{\mu\nu}}{q^+ D}
 -\frac{\alpha}{4\pi^2}
        \langle\gamma^{\mu}\gamma^{\nu}\rangle_{an}
    \frac{g_{\mu\nu}}{p^+ \tilde{D}}
\nonumber\\
 &=& -\frac{\alpha}{4\pi^2}
           \langle\gamma^{\mu}\gamma_{\mu}\rangle_{ex}
         \frac{1}{Q^2}
 -\frac{\alpha}{4\pi^2}
           \langle\gamma^{\mu}\gamma_{\mu}\rangle_{an}
    \frac{1}{M^2}
\,,\label{eq:a1}\end{eqnarray}
where $D=1/2(D_e+D_{\bar{e}})$ and $\tilde{D}=1/2(D_a+D_b)$.
The first choice of similarity function
gives exactly the result of perturbation theory.

\noindent
(2) {\bf Gaussian cut-off}
\begin{eqnarray}
 f(D;l) &=& {\rm exp}\left(- D^2 l \right)
\nonumber\\
 \Theta(D_e,D_{\bar{e}}) &=& \frac{D_e^2}{D_e^2+D_{\bar{e}}^2}
~;~ \vartheta(\xi) = \frac{2\xi}{1+\xi^2} 
\nonumber\\
 V_{\rm eff} &=& \frac{\alpha}{4\pi^2}
           \langle\gamma^{\mu}\gamma^{\nu}\rangle_{ex}
   \left[ \frac{g_{\mu\nu}}{q^+}\frac{D_e+D_{\bar{e}}}{D_e^2+D_{\bar{e}}^2}
 -\frac{\eta_{\mu}\eta_{\nu}}{2{q^+}^2} 
 \frac{(D_e-D_{\bar{e}})^2}{D_e^2+D_{\bar{e}}^2} \right]
\nonumber\\
 &-& \frac{\alpha}{4\pi^2}
           \langle\gamma^{\mu}\gamma^{\nu}\rangle_{an}
\left[ \frac{g_{\mu\nu}}{p^+}\frac{D_a+D_b}{D_a^2+D_b^2}
 -\frac{\eta_{\mu}\eta_{\nu}}{2{p^+}^2}
 \frac{(D_a-D_b)^2}{D_a^2+D_b^2} \right]
\nonumber\\
 = &-& \frac{\alpha}{4\pi^2}
          \langle\gamma^{\mu}\gamma^{\nu}\rangle_{ex}
 \left[ \frac{g_{\mu\nu}}{Q^2}+\frac{\eta_{\mu}\eta_{\nu}}{{q^+}^2}
\frac{\delta Q^4}{Q^4} \right]
\frac{Q^4}{Q^4+\delta Q^4}
\nonumber\\
 &-& \frac{\alpha}{4\pi^2}
           \langle\gamma^{\mu}\gamma^{\nu}\rangle_{an}
 \left[ \frac{g_{\mu\nu}}{M^2}-\frac{\eta_{\mu}\eta_{\nu}}{{p^+}^2}
\frac{\delta M^4}{M^4} \right]
\frac{M^4}{M^4+\delta M^4}
\,,\label{eq:a2}\end{eqnarray}
where we understand under $Q^4= (Q^2)^2$
and $\delta Q^4= (\delta Q^2)^2$ with $Q^2$ and $\delta Q^2$ defined in
Eq.~(\ref{eq:r15}).

\noindent
(3) {\bf Sharp cut-off}
\begin{eqnarray}
 f(D;l) &=& \theta \left(1 -\vert D \vert l \right)
\nonumber\\
 \Theta(D_e,D_{\bar{e}}) &=&
 \theta \left(\vert D_e \vert -\vert D_{\bar{e}} \vert \right)
~;~ \vartheta(\xi) = {\rm sign} (\xi)
\nonumber\\
 V_{\rm eff} &=& \frac{\alpha}{4\pi^2}
        \langle\gamma^{\mu}\gamma^{\nu}\rangle_{ex}
        \left[ \frac{g_{\mu\nu}}{q^+}
 \left( \frac{\theta(\vert D_e\vert-\vert D_{\bar{e}}\vert)}{D_e}
   + \frac{\theta(\vert D_{\bar{e}}\vert-\vert D_e\vert)}{D_{\bar{e}}} 
  \right)  \right.
\nonumber\\
   &&\left. - \frac{\eta_{\mu}\eta_{\nu}}{2{q^+}^2}(D_e-D_{\bar{e}})
 \left( \frac{\theta(\vert D_e\vert-\vert D_{\bar{e}}\vert)}{D_e}
      - \frac{\theta(\vert D_{\bar{e}}\vert-\vert D_e\vert)}{D_{\bar{e}}} 
 \right)    
           \right]
\nonumber\\
 &-& \frac{\alpha}{4\pi^2}
            \langle\gamma^{\mu}\gamma^{\nu}\rangle_{an}
        \left[ \frac{g_{\mu\nu}}{p^+}
 \left( \frac{\theta(\vert D_a\vert-\vert D_b\vert)}{D_a}
      + \frac{\theta(\vert D_b\vert-\vert D_a\vert)}{D_b} \right)\right.
\nonumber\\
   &&\left. - \frac{\eta_{\mu}\eta_{\nu}}{2{p^+}^2}(D_a-D_b)
 \left( \frac{\theta(\vert D_a\vert-\vert D_b\vert)}{D_a}
      - \frac{\theta(\vert D_b\vert-\vert D_a\vert)}{D_b} \right)    
           \right]
\nonumber\\
 = &-& \frac{\alpha}{4\pi^2}
        \langle\gamma^{\mu}\gamma^{\nu}\rangle_{ex}
 \left[ \frac{g_{\mu\nu}}{Q^2}+\frac{\eta_{\mu}\eta_{\nu}}{{q^+}^2}
\frac{\vert \delta Q^2 \vert}{Q^2} \right]
\frac{Q^2}{Q^2+\vert \delta Q^2 \vert}
\nonumber\\
 &-& \frac{\alpha}{4\pi^2}
         \langle\gamma^{\mu}\gamma^{\nu}\rangle_{an}
 \left[ \frac{g_{\mu\nu}}{M^2}-\frac{\eta_{\mu}\eta_{\nu}}{{p^+}^2}
\frac{\vert \delta M^2 \vert}{M^2} \right]
\frac{M^2}{M^2+\vert \delta M^2 \vert}
\,,\label{eq:a3}\end{eqnarray}
The motivation to choose these cut-off functions is the following.
Using exponential cut-off in flow equations one generates
the same interaction as obtained also in Tamm-Dancoff approach,where
numerical calculations of positronium spectrum
are performed \cite{TrPa}, and we use this numerical code here.
Note also, that for this cut-off the effective interaction
looks very much as in covariant calculations: 
it contains only $'g_{\mu\nu}`$ part, and $'\eta_\mu\eta_\nu`$
part is identically zero, so that there is no collinear problem.
Gaussian cut-off corresponds to the original choice of generator
Eqs.(\ref{eq:i5},\ref{eq:ii44}) by Wegner as commutator of diagonal, 
particle number conserving, and off-diagonal, particle number changing, 
parts of Hamiltonian.
Sharp cut-off is used often in the alternative
similarity scheme to perform calculations \cite{GlWi}.

\newpage
\section{The matrix elements in the exchange channel}
\label{app:c}

In this Appendix we follow the scheme of the work \cite{TrPa}
to calculate the matrix elements of the effective interaction
in the exchange channel.\footnote{
Some of these calculations can be found in ~\cite{gub98}. }
Here, we list the general, angle-dependent matrix elements
defining the effective interaction in the exchange channel 
and the corresponding matrix elements
of the effective interaction for arbitrary $J_z$, 
after integrating out the angles.
Exchange part of the effective interaction for three different
cut-offs Eqs.~(\ref{eq:a1}--\ref{eq:a3}) can be written
\begin{eqnarray}
  V_{\rm eff} = - \frac{\alpha}{4\pi^2} 
  \langle\gamma^\mu\gamma^\nu\rangle B_{\mu\nu}   
\,,\end{eqnarray}
where explicitly one has 

\noindent
(1) {\bf Exponential cut-off} 
\begin{eqnarray}
 B_{\mu\nu} = \frac{g_{\mu\nu}}{Q^2}
\,,\label{eq:b1}\end{eqnarray}

\noindent
(2) {\bf Gaussian cut-off}
\begin{eqnarray}
 B_{\mu\nu} = g_{\mu\nu}{\rm Re}\left( \frac{1}{Q^2+i\delta Q^2} \right)
            -\frac{\eta_{\mu}\eta_{\nu}}{{q^+}^2}\delta Q^2
                        {\rm Im}\left( \frac{1}{Q^2+i\delta Q^2} \right)
\,,\label{eq:b2}\end{eqnarray}
where ${\rm Re}$ and ${\rm Im}$ are real and imaginary parts, respectively,
and $i^2=-1$.

\noindent
(3) {\bf Sharp cut-off}
\begin{eqnarray}
  B_{\mu\nu} &=& g_{\mu\nu}
  \left( \frac{\theta(-\delta Q^2)}{Q^2-\delta Q^2}
      +  \frac{\theta( \delta Q^2)}{Q^2+\delta Q^2} \right)
\nonumber\\
             &-& \frac{\eta_{\mu}\eta_{\nu}}{{q^+}^2} \delta Q^2
  \left( \frac{\theta(-\delta Q^2)}{Q^2-\delta Q^2}
      -  \frac{\theta( \delta Q^2)}{Q^2+\delta Q^2} \right)
\,,\label{eq:b3}\end{eqnarray}
where $q=p'_1-p_1$ is the momentum transfer; and
$\langle\gamma^\mu\gamma^\nu\rangle$ for the exchange channel 
is given in Eq.~(\ref{eq:r94}). 
We omit index $`ex'$ everywhere. 

It is convenient to extract the angular dependence in the functions
\begin{eqnarray}
 Q_e^2         &=& a_1 - b\cos t
\nonumber\\
 Q_{\bar{e}}^2 &=& a_2 - b\cos t
\nonumber\\
        t      &=& \varphi-\varphi^{'}
\,,\end{eqnarray}
where we define
 \begin{eqnarray}
 \vec{k}_{\perp} &=& k_{\perp}(\cos\varphi,\sin\varphi)
\end{eqnarray}
in polar system; here the terms are given 
\begin{eqnarray}
 a_1 &=& \frac{x'}{x}k_{\perp}^2+\frac{x}{x'}k_{\perp}^{'2}
+m^2\frac{(x-x')^2}{xx'}
 \nonumber\\
&=& k_{\perp}^2+k_{\perp}^{'2}
 +(x-x')\left(k_{\perp}^2(-\frac{1}{x})-k_{\perp}^{'2}(-\frac{1}{x'})\right)
 +m^2\frac{(x-x')^2}{xx'}
\nonumber\\
 a_2 &=& \frac{1-x'}{1-x}k_{\perp}^2+\frac{1-x}{1-x'}k_{\perp}^{'2}
 +m^2\frac{(x-x')^2}{(1-x)(1-x')}
\nonumber\\
&=& k_{\perp}^2+k_{\perp}^{'2}
 +(x-x')\left(k_{\perp}^2\frac{1}{1-x}-k_{\perp}^{'2}\frac{1}{1-x'}\right)
 +m^2\frac{(x-x')^2}{(1-x)(1-x')}
\nonumber\\
 b &=& 2k_{\perp}k_{\perp}^{'}
\label{eq:b4}\end{eqnarray}
Then the functions in Eqs.~(\ref{eq:b1}--\ref{eq:b3}) are given
\begin{eqnarray}
 Q^2 &=& a - b\cos t
\nonumber\\
 \delta Q^2 &=& \delta a
\,,\end{eqnarray}
where
\begin{eqnarray}
       a &=& \frac{1}{2}(a_1+a_2)
\nonumber\\
\delta a &=& \frac{1}{2}(a_1-a_2)
\,,\end{eqnarray}
It is useful to display the matrix elements of the effective interaction
in the form of tables. The matrix elements depend on the one hand
on the momenta of the electron and positron, respectively, and on the other
hand on their helicities before and after the interaction.
The dependence on the helicities occur during the calculation
of these functions
$E(x,\vec{k}_{\perp};\lambda_1,\lambda_2|x',\vec{k}'_{\perp};
\lambda'_1,\lambda'_2)$
in part I and 
$G(x,k_{\perp};\lambda_1,\lambda_2|x',k'_{\perp};
\lambda'_1,\lambda'_2)$ in part II
as different Kronecker deltas \cite{LeBr}.
These functions are displayed in the form of helicity tables.
We use the following notation for the elements of the tables
\begin{eqnarray}
&& F_i(1,2)~\rightarrow ~E_i(x,\vec{k}_{\perp};x',\vec{k}'_{\perp});~
G_i(x,k_{\perp};x',k'_{\perp})
\label{eq:a4}\end{eqnarray}
Also we have used in both cases for the permutation of particle and
anti-particle
\begin{eqnarray}
&&F_3^{*}(x,\vec{k}_{\perp};x',\vec{k}'_{\perp})
=F_3(1-x,-\vec{k}_{\perp};1-x',-\vec{k}'_{\perp})
\label{eq:a5}\end{eqnarray}
one has the corresponding for the elements of arbitrary $J_z$;
in the case when the function additionally depends
on the component of the total angular momentum $J_z=n$
we have introduced
\begin{eqnarray}
&& \tilde{F}_i(n)=F_i(-n)
\label{eq:a6}\end{eqnarray}

\subsection{The helicity table}

To calculate the matrix elements of the effective interaction
in the exchange channel we use the matrix elements of the Dirac spinors
listed in Table~(\ref{tab:c1}) \cite{LeBr}. 
Also the following holds
$\bar{v}_{\lambda'}(p)\gamma^{\alpha}v_{\lambda}(q)
=\bar{u}_{\lambda}(q)\gamma^{\alpha}u_{\lambda'}(p)$.

\begin{table}[htbp]
\begin{tabular}{|c||c|}
\hline
\parbox{1.5cm}{ \[\cal M\] } & 
\parbox{12cm}{
\[
\frac{1}{\sqrt{k^+k^{'+}}}
\bar{u}(k',\lambda') {\cal M} u(k,\lambda)
\]
}
\\\hline\hline
$\gamma^+$ & \parbox{12cm}{
\[
\hspace{2cm}
2\delta^{\lambda}_{\lambda'}
\]}
\\\hline
$\gamma^-$ &
\parbox{12cm}{
\[
\frac{2}{k^+k^{'+}}\left[\left(m^2+
k_{\perp}k'_{\perp}e^{+i\lambda(\varphi-\varphi')}\right)
\delta^{\lambda}_{\lambda'}
-m\lambda\left(k'_{\perp}e^{+i\lambda \varphi'}-
k_{\perp} e^{+i\lambda\varphi}\right)
\delta^{\lambda}_{-\lambda'}\right]
\]}
\\\hline
$\gamma_{\perp}^1$ & 
\parbox{12cm}{
\[
\left(\frac{k'_{\perp}}{k^{'+}}e^{-i\lambda\varphi'}+
\frac{k_{\perp}}{k^+}e^{+i\lambda\varphi}
\right)\delta^{\lambda}_{\lambda'}
+m\lambda\left(\frac{1}{k^{'+}}-\frac{1}{k^+}
\right)\delta^{\lambda}_{-\lambda'}
\]}
\\\hline
$\gamma_{\perp}^2$ &
\parbox{12cm}{
\[ 
i\lambda\left(\frac{k'_{\perp}}{k^{'+}}e^{-i\lambda\varphi'}-
\frac{k_{\perp}}{k^+}e^{+i\lambda\varphi}
\right)\delta^{\lambda}_{\lambda'}
+im\left(\frac{1}{k^{'+}}-\frac{1}{k^+}
\right)\delta^{\lambda}_{-\lambda'}
\]}\\
\hline
\end{tabular}
\caption{Matrix elements of the Dirac spinors.}
\label{tab:c1}
\end{table}

We introduce for the matrix elements entering in the effective
interaction Eqs.~(\ref{eq:b1}--\ref{eq:b3})
\begin{eqnarray}
   2E^{(1)}(x,\vec{k}_{\perp};\lambda_1,\lambda_2|
   x',\vec{k}_{\perp}^{'};\lambda_1^{'},\lambda_2^{'})
   &=& \langle\gamma^{\mu}\gamma^{\nu}\rangle g_{\mu\nu}
\,,\end{eqnarray}
with $ \langle\gamma^{\mu}\gamma^{\nu}\rangle g_{\mu\nu}
   = \frac{1}{2}\langle\gamma^+\gamma^-\rangle
   + \frac{1}{2}\langle\gamma^-\gamma^+\rangle
   - \langle\gamma_1^2\rangle-\langle\gamma_2^2\rangle$ 
and
\begin{eqnarray}
   2E^{(2)}(x,\vec{k}_{\perp};\lambda_1,\lambda_2|
   x',\vec{k}_{\perp}^{'};\lambda_1^{'},\lambda_2^{'})
   &=& \langle\gamma^{\mu}\gamma^{\nu}\rangle 
     \eta_{\mu}\eta_{\nu}\frac{1}{{q^+}^2} 
\,,\end{eqnarray}
with $\langle\gamma^{\mu}\gamma^{\nu}\rangle 
    \eta_{\mu}\eta_{\nu}
   = \langle\gamma^{+}\gamma^{+}\rangle $; 
where
\begin{eqnarray}
   \langle\gamma^{\mu}\gamma^{\nu}\rangle = 
   {(\bar{u}(x,\vec{k}_{\perp};\lambda_1)\gamma^{\mu}
   u(x',\vec{k}_{\perp}^{'};\lambda_1^{'}))
   \over \sqrt{xx'}}
   {(\bar{v}(1-x',-\vec{k}_{\perp}^{'};\lambda_2^{'})
   \,\gamma^{\nu}
   v(1-x,-\vec{k}_{\perp};\lambda_2))
   \over \sqrt{(1-x)(1-x')}}
\label{eq:a8}\end{eqnarray}

These functions are displayed in
Table~(\ref{tab:c2}).

\begin{table}[htb]
\begin{tabular}{|c||c|c|c|c|}\hline
\rule[-3mm]{0mm}{8mm}{\bf final : initial} & 
$(\lambda_1',\lambda_2')=\uparrow\uparrow$ 
& $(\lambda_1',\lambda_2')=\uparrow\downarrow$ 
& $(\lambda_1',\lambda_2')=\downarrow\uparrow$ &
$(\lambda_1',\lambda_2')=\downarrow\downarrow$ \\ \hline\hline
\rule[-3mm]{0mm}{8mm}$(\lambda_1,\lambda_2)=\uparrow\uparrow$ & $E_1(1,2)$  
& $E_3^*(1,2)$ & $E_3(1,2)$ & $0$ \\ \hline
\rule[-3mm]{0mm}{8mm}$(\lambda_1,\lambda_2)=\uparrow\downarrow$ & 
$E_3^*(2,1)$ & $E_2(1,2)$ & $E_4(1,2)$ 
& $-E_3(2,1)$ \\ \hline
\rule[-3mm]{0mm}{8mm}$(\lambda_1,\lambda_2)=\downarrow\uparrow$& $E_3(2,1)$ 
& $E_4(1,2)$ & $E_2(1,2)$  & 
$-E_3^*(2,1)$\\ \hline
\rule[-3mm]{0mm}{8mm}$(\lambda_1,\lambda_2)=\downarrow\downarrow$ & $0$ 
& $-E_3(1,2)$ & $-E_3^*(1,2)$ & 
$E_1(1,2)$\\ \hline
\end{tabular}
\caption{General helicity table defining the effective interaction
in the exchange channel.}
\label{tab:c2}
\end{table}

The matrix elements 
$E_i^{(n)}(1,2)=E_i^{(n)}(x,\vec{k}_{\perp};x',\vec{k}'_{\perp})$
with $n=1$ and $n=2$ for $`g_{\mu\nu}'$ and $\eta_{\mu}\eta_{\nu}$ terms, 
respectively, are the following
\begin{eqnarray}
  E_1^{(1)}(x,\vec{k}_{\perp};x',\vec{k}_{\perp}^{'})
&=& m^2\left(\frac{1}{xx'}+\frac{1}{(1-x)(1-x')}\right)
    +\frac{k_{\perp}k_{\perp}^{'}}{xx'(1-x)(1-x')}
    {\rm e}^{-i(\varphi-\varphi^{'})}\nonumber\\
  E_2^{(1)}(x,\vec{k}_{\perp};x',\vec{k}_{\perp}^{'})
&=& m^2\left(\frac{1}{xx'}+\frac{1}{(1-x)(1-x')}\right)
    +k_{\perp}^2\frac{1}{x(1-x)}+k_{\perp}^{'2}\frac{1}{x'(1-x')}\nonumber\\
&+& k_{\perp}k_{\perp}^{'}
    \left(\frac{{\rm e}^{i(\varphi-\varphi^{'})}}{xx'}
    +\frac{{\rm e}^{-i(\varphi-\varphi^{'})}}{(1-x)(1-x')}\right)\nonumber\\
  E_3^{(1)}(x,\vec{k}_{\perp};x',\vec{k}_{\perp}^{'})
&=& -\frac{m}{xx'}
    \left(k_{\perp}^{'}{\rm e}^{i\varphi^{'}}
    -k_{\perp}\frac{1-x'}{1-x}{\rm e}^{i\varphi}\right)\nonumber\\
  E_4^{(1)}(x,\vec{k}_{\perp};x',\vec{k}_{\perp}^{'})
&=& -m^2\frac{(x-x')^2}{xx'(1-x)(1-x')}
\,,\label{eq:a9}\end{eqnarray}
and
\begin{eqnarray}
  E_1^{(2)}(x,\vec{k}_{\perp};x',\vec{k}_{\perp}^{'})
&=& E_2^{(2)}(x,\vec{k}_{\perp};x',\vec{k}_{\perp}^{'})
 =  \frac{2}{(x-x')^2}
\nonumber\\
  E_3^{(2)}(x,\vec{k}_{\perp};x',\vec{k}_{\perp}^{'})
&=& E_4^{(2)}(x,\vec{k}_{\perp};x',\vec{k}_{\perp}^{'})
 =  0
\,.\label{eq:ab9}\end{eqnarray}

\subsection{The helicity table for arbitrary $J_z$.}

Following the description given in the main text Eq.~(\ref{eq:r43})
we integrate out the angles in the effective interaction 
in the exchange channel.
For the matrix elements of the effective interaction
for an arbitrary $J_z=n$ with $n\in {\bf Z}$ we introduce the functions
$G(x,k_{\perp};\lambda_1,\lambda_2|x',k'_{\perp};\lambda'_1,\lambda'_2)=
\langle x,k_{\perp};J_z,\lambda_1,\lambda_2|\tilde{V}_{\rm eff}|
x',k'_{\perp};J'_z,\lambda'_1,\lambda'_2\rangle$ 
in the exchange channel
and obtain the helicity Table~(\ref{tab:c3}).

\begin{table}[htb]
\begin{tabular}{|c||c|c|c|c|}\hline
\rule[-3mm]{0mm}{8mm}{\bf final : initial} 
&$(\lambda'_1,\lambda'_2)=\uparrow\uparrow$
&$(\lambda'_1,\lambda'_2)=\uparrow\downarrow$
&$(\lambda'_1,\lambda'_2)=\downarrow\uparrow$
&$(\lambda'_1,\lambda'_2)=\downarrow\downarrow$\\\hline\hline
\rule[-3mm]{0mm}{8mm}$(\lambda_1,\lambda_2)=\uparrow\uparrow$
&$G_1(1,2)$&$G_3^*(1,2)$&$G_3(1,2)$&$0$\\\hline
\rule[-3mm]{0mm}{8mm}$(\lambda_1,\lambda_2)=\uparrow\downarrow$
&$G_3^*(2,1)$&$G_2(1,2)$&$G_4(1,2)$&$-\tilde{G}_3(2,1)$\\\hline
\rule[-3mm]{0mm}{8mm}$(\lambda_1,\lambda_2)=\downarrow\uparrow$
&$G_3(2,1)$&$G_4(1,2)$&$\tilde{G}_2(1,2)$&$-\tilde{G}_3^*(2,1)$\\\hline
\rule[-3mm]{0mm}{8mm}$(\lambda_1,\lambda_2)=\downarrow\downarrow$&$0$&$-
\tilde{G}_3(1,2)$&$-\tilde{G}_3^*(1,2)$&
$\tilde{G}_1(1,2)$\\
\hline
\end{tabular}
\caption{Helicity table of the effective interaction
for $J_z = \pm n$, $x>x'$.}
\label{tab:c3}
\end{table}

Here, the functions $G_i(1,2)=G_i(x,k_{\perp};x',k'_{\perp})$
are given
\begin{eqnarray}
 G_1(x,k_{\perp};x',k_{\perp}^{'}) &=& 
    \left( \frac{m^2}{xx'}+\frac{m^2}{(1-x)(1-x')} \right)Int(|1-n|)
\nonumber\\
&+& \frac{k_{\perp}k_{\perp}^{'}}{xx'(1-x)(1-x')}Int(|n|)
 -  \frac{2\delta a}{(x-x')^2}\tilde{Int}(|1-n|)
\nonumber\\
 G_2(x,k_{\perp};x',k_{\perp}^{'}) &=& \left(
    m^2\left(\frac{1}{xx'}+\frac{1}{(1-x)(1-x')} \right)
 +  \frac{k_{\perp}^2}{x(1-x)}+\frac{k_{\perp}^{'2}}{x'(1-x')} 
    \right)Int(|n|)
\nonumber\\
&+& k_{\perp}k_{\perp}^{'}\left( \frac{1}{xx'}Int(|1-n|)
 +  \frac{1}{(1-x)(1-x')}Int(|1+n|) \right)
\nonumber\\
&-& \frac{2\delta a}{(x-x')^2}\tilde{Int}(|n|)
\nonumber\\
 G_3(x,k_{\perp};x',k_{\perp}^{'}) &=&
 -  \frac{m}{xx'}\left(
    k_{\perp}^{'}Int(|1-n|)
 -  k_{\perp}\frac{1-x'}{1-x}Int(|n|) 
    \right)
\nonumber\\
 G_4(x,k_{\perp};x',k_{\perp}^{'}) &=&
 -  m^2\frac{(x-x')^2}{xx'(1-x)(1-x')}Int(|n|)
\label{eq:a11}\end{eqnarray}
we define
\begin{eqnarray}
 I(n;a,b) = -\frac{\alpha}{2\pi^2} 
 \int_0^{2\pi}dt\frac{\cos nt}{a-b\cos t}
\,,\end{eqnarray}
then in Eq.~(\ref{eq:a11}) the following functions are introduced

\noindent
(1) {\bf Exponential cut-off} 
\begin{eqnarray}
 Int(n) &=& I(n;a,b)
\nonumber\\
 \tilde{Int}(n) &=& 0
\,,\label{eq:d1}\end{eqnarray}

\noindent
(2) {\bf Gaussian cut-off}
\begin{eqnarray}
 Int(n)         &=& {\rm Re}I(n;a+i\delta a,b)
\nonumber\\
 \tilde{Int}(n) &=& {\rm Im}I(n;a+i\delta a,b)
\,,\label{eq:d2}\end{eqnarray}

\noindent
(3) {\bf Sharp cut-off}
\begin{eqnarray}
         Int(n) &=& \theta(-\delta a)I(n;a-\delta a,b)
                 +  \theta( \delta a)I(n;a+\delta a,b)
\nonumber\\
 \tilde{Int}(n) &=& \theta(-\delta a)I(n;a-\delta a,b)
                 -  \theta( \delta a)I(n;a+\delta a,b)
\,,\label{eq:d3}\end{eqnarray}
also $a+\delta a =a_1$ and$a-\delta a =a_2$.  

Explicitly is used 
\begin{eqnarray}
 && \int_0^{2\pi}dt\frac{\cos nt}{a-b\cos t}
 =  2\pi\frac{1}{\sqrt{a^2-b^2}}
    \left( \frac{a-\sqrt{a^2-b^2}}{b} \right)^n
\nonumber\\
 && \int_0^{2\pi}dt\frac{\sin nt}{a-b\cos t}
 =  0
\,,\label{eq:int}\end{eqnarray}
where $a$ can contain imaginary part 
as in the case of gaussian cut-off.
%

\newpage
\section{The matrix elements in the annihilation channel}
\label{app:d}

We repeat the same calculations for the matrix elements 
of the effective interaction in the annihilation channel. 
Annihilation part of the effective interaction 
can be written
\begin{eqnarray}
  V_{\rm eff} = - \frac{\alpha}{4\pi^2} 
  \langle\gamma^\mu\gamma^\nu\rangle C_{\mu\nu}   
\,,\end{eqnarray}
where one has
\begin{eqnarray}
   C_{\mu\nu} &=& 
   g_{\mu\nu}^{\perp}
   \left(\frac{\Theta_{ab}} {M_a^2} +
         \frac{\Theta_{ba}} {M_b^2} \right) - 
   \frac{\eta_{\mu}\eta_{\nu}}{{p^+}^2}
\nonumber\\
              &=& \frac{g_{\mu\nu}^{\perp}}{M^2}\
   \frac{1-\beta\chi(\beta)}{1-\beta^2}
  -\frac{\eta_{\mu}\eta_{\nu}}{{p^+}^2}
\,,\end{eqnarray}
in the frame $p_{\perp}=0$.\footnote{
Indeed
$\langle\gamma^\mu\gamma^\nu\rangle g_{\mu\nu}
=\frac{1}{2}\langle\gamma^+\gamma^-\rangle
+\frac{1}{2}\langle\gamma^-\gamma^+\rangle 
+\langle\gamma^\mu\gamma^\nu\rangle 
g_{\mu\nu}^{\perp}$;
therefore it holds
\begin{eqnarray}
  g_{\mu\nu} = g_{\mu\nu}^{\perp}
+ \frac{\eta_\mu(p_\nu-p_\nu^{\perp})+\eta_\nu(p_\mu-p_\mu^{\perp})}{p^+}
- \frac{{p^{\perp}}^2}{{p^+}^2}\eta_\mu\eta_\nu
\nonumber 
\,,\end{eqnarray}
The $4$-momentum of the photon $p_{\mu}$ in the $t$-channel can be written
$p_{\mu}=p^{'}_{1\mu}+p^{'}_{2\mu}-\eta_{\mu}D_a/2
        =p_{1\mu}    +p_{2\mu}    -\eta_{\mu}D_b/2$ with $D_a,D_b$ defined
in Eq.~(\ref{eq:40b}).
The Dirac equation $(p_1+p_2)_{\mu}\bar{u}(p_1)\gamma^{\mu}v(p_2)=0$ 
allows then to write 
$p_{\mu}\bar{u}(p_1,\lambda_1)\gamma^{\mu}v(p_2,\lambda_2)
 =   -M_b^2/(2p^+)\eta_{\mu}              
        \bar{u}(p_1,\lambda_1)\gamma^{\mu}v(p_2,\lambda_2)$.
Thus, when $p_{\perp}=0$, one has
\begin{eqnarray} 
g_{\mu\nu} \rightarrow g_{\mu\nu}^{\perp}-\frac{\eta_\mu\eta_\nu}{{p^+}^2}M^2
\nonumber
\,,\end{eqnarray}
where the arrow means that this tensor should be contracted with 
$\langle\gamma^\mu\gamma^\nu\rangle$ in the annihilation channel.
}
Explicitly the annihilation part of the effective interaction
for different cut-offs Eq.~({\ref{eq:a1})- Eq.~(\ref{eq:a3}) is given

\noindent
(1) {\bf Exponential cut-off} 
\begin{eqnarray}
 C_{\mu\nu} = \frac{g_{\mu\nu}^{\perp}}{M^2}
\,,\label{eq:f1}\end{eqnarray}

\noindent
(2) {\bf Gaussian cut-off}
\begin{eqnarray}
 C_{\mu\nu} = g_{\mu\nu}^{\perp}\frac{M^2}{M^4+\delta M^4}
            - \frac{\eta_{\mu}\eta_{\nu}}{{p^+}^2}
\,,\label{eq:f2}\end{eqnarray}
             
\noindent
(3) {\bf Sharp cut-off}
\begin{eqnarray}
  C_{\mu\nu} = g_{\mu\nu}^{\perp}\left( 
               \frac{\theta(M_a^2-M_b^2)}{M_a^2}
             + \frac{\theta(M_b^2-M_a^2)}{M_b^2} \right)       
             - \frac{\eta_{\mu}\eta_{\nu}}{{p^+}^2}
\,\label{eq:f3}\end{eqnarray}
where
$p^+=p_1^++p_2^+$ is the total momentum; and 
$\langle\gamma^\mu\gamma^\nu\rangle$
for annihilation is defined in Eq.~(\ref{eq:r94}).
The functions present in Eq.~(\ref{eq:f1})- Eq.~(\ref{eq:f3}) 
are given in the light-front frame 
\begin{eqnarray}
 M_a^2 &=& \frac{{k'_{\perp}}^2+m^2}{x'(1-x')}
\nonumber\\
 M_b^2 &=&  \frac{k_{\perp}^2+m^2}{x(1-x)}
\,,\label{eq:a12}\end{eqnarray}
we remind also
\begin{eqnarray}
 M^2 &=& \frac{1}{2}(M_a^2+M_b^2)
\nonumber\\
 \delta M^2 &=& \frac{1}{2}(M_a^2-M_b^2)
\label{eq:a13}\end{eqnarray}

Note that the energy denominators of the effective interaction
in the annihilation channel 
do not depend on the angles $\varphi,\varphi'$.

\begin{table}[htb]
\begin{tabular}{|c||c|}
\hline
\parbox{1.5cm}{ \[\cal M\] } & 
\parbox{12.5cm}{
\[
\frac{1}{\sqrt{k^+k^{'+}}}
\bar{v}(k',\lambda') {\cal M} u(k,\lambda)
\]
}
\\\hline\hline
$\gamma^+$ & \parbox{12.5cm}{
\[
\hspace{2cm}
2\delta^{\lambda}_{-\lambda'}
\]}
\\\hline
$\gamma^-$ &
\parbox{12.5cm}{
\[
\frac{2}{k^+k^{'+}}\left[-\left(m^2-
k_{\perp}k'_{\perp}e^{+i\lambda(\varphi-\varphi')}\right)
\delta^{\lambda}_{-\lambda'}
-m\lambda\left(k'_{\perp}e^{+i\lambda \varphi'}+
k_{\perp} e^{+i\lambda\varphi}\right)
\delta^{\lambda}_{\lambda'}\right]
\]}
\\\hline
$\gamma_{\perp}^1$ & 
\parbox{12.5cm}{
\[
\left(\frac{k'_{\perp}}{k^{'+}}e^{-i\lambda\varphi'}+
\frac{k_{\perp}}{k^+}e^{+i\lambda\varphi}
\right)\delta^{\lambda}_{-\lambda'}
-m\lambda\left(\frac{1}{k^{'+}}+\frac{1}{k^+}
\right)\delta^{\lambda}_{\lambda'}
\]}
\\\hline
$\gamma_{\perp}^2$ &
\parbox{12.5cm}{
\[ 
i\lambda\left(\frac{k'_{\perp}}{k^{'+}}e^{-i\lambda\varphi'}-
\frac{k_{\perp}}{k^+}e^{+i\lambda\varphi}
\right)\delta^{\lambda}_{-\lambda'}
-im\left(\frac{1}{k^{'+}}+\frac{1}{k^+}
\right)\delta^{\lambda}_{\lambda'}
\]}\\
\hline
\end{tabular}
\caption{Matrix elements of the Dirac spinors.}
\label{tab:d1}
\end{table}

\subsection{The helicity table}

For the calculation of matrix elements of effective interaction
in the annihilation channel 
we use the matrix elements of the Dirac spinors 
listed in Table~(\ref{tab:d1}) \cite{LeBr}.  
Also the following holds
$ (\bar{v}_{\lambda'}(p)\gamma^{\alpha}u_{\lambda}(q))^{+}
 = \bar{u}_{\lambda}(q)\gamma^{\alpha}v_{\lambda'}(p)$.

\noindent
We introduce
\begin{eqnarray}
 2H^{(1)}(x,\vec{k}_{\perp};\lambda_1,\lambda_2|x',\vec{k}_{\perp}^{'};
 \lambda'_1,\lambda'_2)
 &=& \langle\gamma^{\mu}\gamma^{\nu}\rangle g_{\mu\nu}^{\perp}
  = -\langle\gamma_1^2\rangle-\langle\gamma_2^2\rangle
\nonumber\\
 2H^{(2)}(x,\vec{k}_{\perp};\lambda_1,\lambda_2|x',\vec{k}_{\perp}^{'};
 \lambda'_1,\lambda'_2)
 &=& \langle\gamma^{\mu}\gamma^{\nu}\rangle 
  \eta_{\mu}\eta_{\nu}\frac{1}{p^{+2}}
\label{eq:a16}\end{eqnarray}
where
\begin{eqnarray}
  \langle\gamma^{\mu}\gamma^{\nu}\rangle
= \frac{(\bar{v}(1-x',-\vec{k}_{\perp}^{'};\lambda'_2)\gamma^{\mu}
  u(x',\vec{k}_{\perp}^{'};\lambda'_1))}{\sqrt{x'(1-x')}}
  \frac{(\bar{u}(x,\vec{k}_{\perp};\lambda_1) \gamma^{\nu}
  v(1-x,-\vec{k}_{\perp};\lambda_2)}{\sqrt{x(1-x)}}
\label{eq:a14}\end{eqnarray}
These functions are displayed in the Table~(\ref{tab:d2}).

\begin{table}[htb]
\begin{tabular}{|c||c|c|c|c|}\hline
\rule[-3mm]{0mm}{8mm}{\bf final:initial} & $(\lambda'_1,\lambda'_2)
=\uparrow\uparrow$ 
& $(\lambda'_1,\lambda'_2)=\uparrow\downarrow$ 
& $(\lambda'_1,\lambda'_2)=\downarrow\uparrow$ &
$(\lambda'_1,\lambda'_2)=\downarrow\downarrow$ \\ \hline\hline
\rule[-3mm]{0mm}{8mm}$(\lambda_1,\lambda_2)=\uparrow\uparrow$ & 
$H_1(1,2)$   
&$H_3(2,1)$ & $H^*_3(2,1)$ & $0$ \\ \hline
\rule[-3mm]{0mm}{8mm}$(\lambda_1,\lambda_2)=\uparrow\downarrow$ & 
$H_3(1,2)$ 
& $H^*_2(1,2)$ & $H_4(2,1)$ &$0$ \\ \hline
\rule[-3mm]{0mm}{8mm}$(\lambda_1,\lambda_2)=\downarrow\uparrow$& 
$H_3^*(1,2)$ & $H_4(1,2)$ & $H_2(1,2)$  & $0$\\ \hline
\rule[-3mm]{0mm}{8mm} $(\lambda_1,\lambda_2)=\downarrow\downarrow$ & $0$ 
& $0$ & $0$ & $0$  \\
\hline
\end{tabular}
\caption{General helicity table 
defining the effective interaction in the annihilation channel.}
\label{tab:d2}
\end{table}
\vspace{0.5cm}

Here, the matrix elements 
$H^{(n)}_i(1,2)=H^{(n)}_i(x,\vec{k}_{\perp};x',\vec{k}_{\perp}^{'})$
are the following
\begin{eqnarray}
    H^{(1)}_1(x,\vec{k}_{\perp};x',\vec{k}_{\perp}^{'})
&=& - m^2\left(\frac{1}{x}+\frac{1}{1-x}\right)
         \left(\frac{1}{x'}+\frac{1}{1-x'}\right)
\nonumber\\
    H^{(1)}_2(x,\vec{k}_{\perp};x',\vec{k}_{\perp}^{'})
&=& - k_{\perp}k'_{\perp}\left( \frac{{\rm e}^{ i(\varphi-\varphi^{'})}}{xx'} \right)
\nonumber\\
    H^{(1)}_3(x,\vec{k}_{\perp};x',\vec{k}_{\perp}^{'})
&=& - m\lambda_1\left(\frac{1}{x}+\frac{1}{1-x}\right)
                 \frac{k_{\perp}^{'}}{1-x'}{\rm e}^{i\varphi}
\nonumber\\
    H^{(1)}_4(x,\vec{k}_{\perp};x',\vec{k}_{\perp}^{'})
&=&   k_{\perp}k'_{\perp}
           \left( \frac{{\rm e}^{i(\varphi-\varphi^{'})}}{x'(1-x)} \right)
\label{eq:a17}\end{eqnarray}

\noindent
and
\begin{eqnarray}
&& H^{(2)}_1(x,\vec{k}_{\perp};x',\vec{k}_{\perp}^{'})
=H^{(2)}_3(x,\vec{k}_{\perp};x',\vec{k}_{\perp}^{'})=0
\nonumber\\
&& H^{(2)}_2(x,\vec{k}_{\perp};x',\vec{k}_{\perp}^{'})
=H^{(2)}_4(x,\vec{k}_{\perp};x',\vec{k}_{\perp}^{'})=2
\label{eq:a15}\end{eqnarray}

\subsection{The helicity table for $|J_z|\leq 1$}

The matrix elements of the effective interaction
for $J_z\geq 0$~~ 
$F(x,k_{\perp};\lambda_1,\lambda_2|x',k'_{\perp};
\lambda'_1,\lambda'_2)=
\langle x,k_{\perp};J_z,\lambda_1,\lambda_2|\tilde{V}_{eff}|
x',k'_{\perp};J'_z,\lambda'_1,\lambda'_2\rangle$ in the annihilation channel
(the sum of the generated interaction for $J_z=+1$
and instantaneous graph for $J_z=0$)
are given in Table~(\ref{tab:d3}).

\begin{table}[htb]
\begin{tabular}{|c||c|c|c|c|}\hline
\rule[-3mm]{0mm}{8mm}{\bf final:initial} & $(\lambda'_1,\lambda'_2)
=\uparrow\uparrow$ 
& $(\lambda'_1,\lambda'_2)=\uparrow\downarrow$ 
& $(\lambda'_1,\lambda'_2)=\downarrow\uparrow$ &
$(\lambda'_1,\lambda'_2)=\downarrow\downarrow$ \\ \hline\hline
\rule[-3mm]{0mm}{8mm}$(\lambda_1,\lambda_2)=\uparrow\uparrow$ & 
$F_1(1,2)$   
&$F_3(2,1)$ & $F^*_3(2,1)$ & $0$ \\ \hline
\rule[-3mm]{0mm}{8mm}$(\lambda_1,\lambda_2)=\uparrow\downarrow$ & 
$F_3(1,2)$ 
& $F^*_2(1,2)$ & $F_4(2,1)$ &$0$ \\ \hline
\rule[-3mm]{0mm}{8mm}$(\lambda_1,\lambda_2)=\downarrow\uparrow$& 
$F_3^*(1,2)$ & $F_4(1,2)$ & $F_2(1,2)$  & $0$\\ \hline
\rule[-3mm]{0mm}{8mm} $(\lambda_1,\lambda_2)=\downarrow\downarrow$ & $0$ 
& $0$ & $0$ & $0$  \\
\hline
\end{tabular}
\caption{Helicity table of the effective interaction
in the annihilation channel for $J_z\ge 0$.}
\label{tab:d3}
\end{table}
\vspace{0.5cm}

The function $F_i(1,2)=F_i(x,k_{\perp};x',k'_{\perp})$ are 
the following
\begin{eqnarray}
  F_1(x,k_{\perp};x',k_{\perp}^{'})
&=& \frac{\alpha}{\pi}\frac{1}{\Omega}
    \frac{m^2}{xx'(1-x)(1-x')}\delta_{|J_z|,1}
\nonumber\\
  F_2(x,k_{\perp};x',k_{\perp}^{'})
&=& \frac{\alpha}{\pi}\left(\frac{1}{\Omega}
    \frac{k_{\perp}k'_{\perp}}{xx'}\delta_{|J_z|,1}
    +2\delta_{J_z,0}\right)
\nonumber\\
  F_3(x,k_{\perp};x',k_{\perp}^{'})
&=& \frac{\alpha}{\pi}\frac{1}{\Omega}
    \lambda_1\frac{m}{x'(1-x')}
    \frac{k_{\perp}}{1-x}\delta_{|J_z|,1}
\nonumber\\
  F_4(x,k_{\perp};x',k_{\perp}^{'})
&=& \frac{\alpha}{\pi}\left(-\frac{1}{\Omega}
    \frac{k_{\perp}k'_{\perp}}{x(1-x')}\delta_{|J_z|,1}
    +2\delta_{J_z,0}\right)
\label{eq:a12a}\end{eqnarray}
where we have introduced

\noindent
(1) {\bf Exponential cut-off} 
\begin{eqnarray}
 \frac{1}{\Omega} = \frac{1}{M^2}
\,,\end{eqnarray}

\noindent
(2) {\bf Gaussian cut-off}
\begin{eqnarray}
 \frac{1}{\Omega} = \frac{M_a^2+M_b^2}{M_a^4 + M_b^4}
\,,\end{eqnarray}

\noindent
(3) {\bf Sharp cut-off}
\begin{eqnarray}
\frac{1}{\Omega} &=& \frac{\theta(M_a^2-M_b^2)}{M_a^2}
                  +  \frac{\theta(M_b^2-M_a^2)}{M_b^2}
\,.\end{eqnarray}

The table for $J_z=-1$ is obtained by inverting all helicities, i.e.
\begin{eqnarray}
&& F(J_z=+1;\lambda_1,\lambda_2)=-\lambda_1 F(J_z=-1;-\lambda_1,-\lambda_2)
\,,\label{eq:a18}\end{eqnarray}

The matrix elements of the effective interaction 
in the annihilation channel are nonzero only for $|J_z|\leq 1$
due to the restriction on the angular momentum of the photon.

\end{document}

%% file: fig12.tex
\begin{figure}
\unitlength1cm   
\begin{minipage}[t]{60mm}
\epsfxsize=60mm\epsfbox{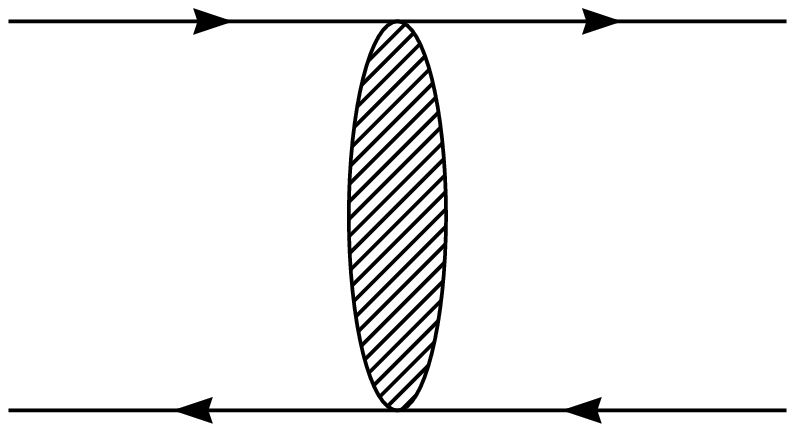}
\caption{\label{fig:1}\em
      The effective interaction between an electron ($e$)
      and a positron ($\bar e$).
}\end{minipage}
\hfill
\begin{minipage}[t]{60mm}
\epsfxsize=60mm\epsfbox{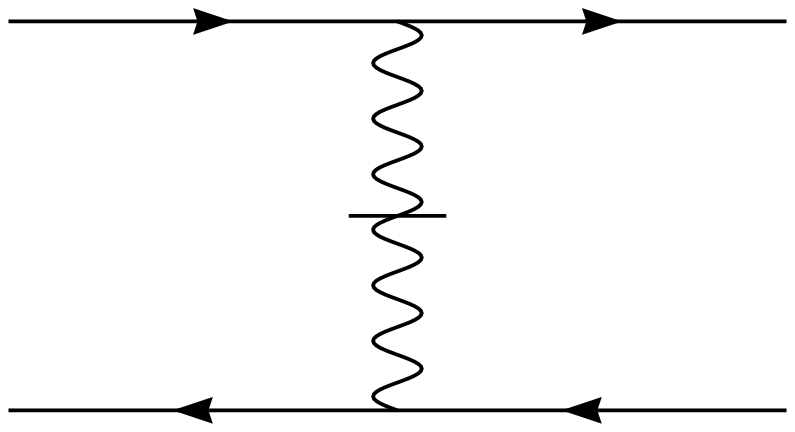}
\caption{\em 
      The graph of the instantaneous exchange interaction.
      Taken from \protect{\cite{gub98}}.
}\end{minipage}
\end{figure} 